\documentclass[prd,aps,showpacs,floats,letterpaper,floatfix,twocolumn,
nofootinbib]{revtex4}
\usepackage{amssymb,amsmath}
\usepackage[dvips]{graphicx}
\usepackage{hyperref}

\newcommand{\mylistitem}[1]{\vspace{6pt}\noindent\textbf{#1}}
\newcommand{\lastlistitem}{\vspace{6pt}}
\newcommand{\doublemylistitem}[1]{\vspace{12pt}\noindent\textbf{#1}}

\newcommand{\moverrightarrow}[1]{\overrightarrow{\vphantom{3'}#1}}
\newcommand{\moverleftarrow}[1]{\overleftarrow{\vphantom{3'}#1}}

\begin{document}

\title{Geometric Time Delay Interferometry}

\author{Michele Vallisneri}

\affiliation{Jet Propulsion Laboratory, California Institute of
Technology, Pasadena, CA 91109}

\begin{abstract}
The space-based gravitational-wave observatory LISA, a NASA--ESA mission
to be launched after 2012, will achieve its optimal sensitivity using
Time Delay Interferometry (TDI), a LISA-specific technique needed to
cancel the otherwise overwhelming laser noise in the inter-spacecraft
phase measurements. The TDI observables of the \emph{Michelson} and \emph{Sagnac} types have been interpreted physically as the virtual measurements of a synthesized interferometer. In this paper, I present \emph{Geometric TDI}, a new and intuitive approach to extend this interpretation to \emph{all} TDI observables. Unlike the standard algebraic formalism, Geometric TDI provides a
combinatorial algorithm to explore exhaustively the space of
\emph{second-generation} TDI observables (i.e., those that cancel laser
noise in LISA-like interferometers with time-dependent armlengths).
Using this algorithm, I survey the space of second-generation TDI
observables of length (i.e., number of component phase measurements) up
to 24, and I identify alternative, improved forms of the standard
second-generation TDI observables. The alternative forms have improved
high-frequency gravitational-wave sensitivity in realistic noise
conditions (because they have fewer nulls in the gravitational-wave and
noise response functions), and are less susceptible to instrumental gaps
and glitches (because their component phase measurements span shorter
time periods).
\end{abstract}

\date{Version of June 28, 2005}


\maketitle

\section{Introduction}
 
The Laser Interferometer Space Antenna (LISA) is a joint NASA--ESA
deep-space mission to be launched after 2012, aimed at detecting and
studying gravitational waves (GWs) with frequencies between $10^{-5}$
and $10^{-1}$ Hz \cite{PPA98}. LISA will provide access to GW sources
that are outside the reach of ground-based interferometric GW detectors
\cite{ligovirgo}, such as the binaries of compact stellar objects in our
galaxy, the mergers of massive and supermassive black holes, and the
gravitational captures of compact objects by the supermassive black
holes at the center of galaxies \cite{review}.

LISA consists of three widely separated spacecraft, flying around the
Sun in a quasi-equilateral triangular configuration and exchanging
phase-coherent laser signals. LISA relies on picometer interferometry to
measure GWs as modulations in the distance between the spacecraft. The
greatest challenge to achieving this measurement is the phase noise of
the LISA lasers, which is larger than the GW-induced response by many
orders of magnitude, and which cannot be removed by conventional
phase-matching interferometry because the LISA armlengths are grossly
unequal, and changing continuously. \emph{Time Delay Interferometry}
(TDI), developed by J. W. Armstrong, F. B. Estabrook, M. Tinto, and
others \cite{tintoarm,firstgen,firstsens,dhurandhar,shaddock2004,cornhell,stea2003,tea2004}, is the LISA-specific technique that will be used to
combine the laser-noise--laden one-way phase measurements performed
between the three spacecraft\footnote{A variant of the technique uses
combinations of one-way and \emph{two-way} phase measurements, generated
by locking five of the six LISA lasers to the last one, as described by
Tinto and colleagues \cite{TSSA2003}. In this paper we shall consider
only the one-way formalism, but our results could be applied with
superficial modifications also to the two-way variant.} into virtual
interferometric observables where laser noise is reduced by several
orders of magnitude.

TDI was initially developed using \emph{ad hoc} algebraic reasoning for
the case of a stationary LISA configuration with unequal but constant
armlengths (\emph{first-generation} TDI, see \cite{tintoarm,firstgen}).
It was later modified to work also in the case of a rotating LISA
constellation (\emph{modified} TDI, see \cite{shaddock2004,cornhell,stea2003,tea2004})
and of linearly changing armlengths (\emph{second-generation} TDI, see
\cite{stea2003,tea2004}). First-generation and modified TDI were given a rigorous mathematical foundation in the theory of algebraic syzygies on moduli \cite{dhurandhar}, providing tools to generate all possible TDI
observables, and to determine which observables are optimally sensitive
to GWs \cite{optimal}. Unfortunately, this algebraic treatment cannot be
extended easily to second-generation TDI, which is the version that must
be used in practice.

In this paper, I give a new derivation of first-generation, modified,
and second-generation TDI, using a \emph{geometric} approach that
emphasizes the physical interpretation of TDI observables as synthesized
interferometric measurements \cite{tintoarm,stea2003,summers}, extending it
to all known observables. What is more, this geometric approach to
TDI (in short, \emph{Geometric TDI}) allows the exhaustive enumeration
of all TDI observables of any length, and it leads to alternative,
improved forms of the standard TDI observables, characterized by better
GW sensitivity at high frequencies in realistic noise conditions, by
lesser demands on the measurement system, and by reduced susceptibility
to gaps and glitches.

More specifically, all TDI observables display nulls in their noise and
GW responses at frequency multiples of the inverse arm-crossing light
times.\footnote{The responses are exactly null only in the limit of
equal LISA armlengths. For realistic, time-evolving LISA geometries the
nulls are spread into narrow dips; however, these are deep enough that
the qualitative discussion to follow still applies.} Because these zeros
occur at the same frequencies and with the same orders for noise and
GWs, the \emph{ideal} GW sensitivity after successful laser-noise
suppression is finite, and comparable to the sensitivity at nearby
frequencies. The \emph{actual} sensitivity, however, is likely to be
degraded, either because noise \emph{leaks} into the nulls from the
sides \cite{jenkinswatts}, or because the measurement system has
insufficient dynamical range to resolve the tiny signals within the
nulls. This problem is mitigated with the alternative observables, which
have half as many response-function nulls as the standard forms.

In addition, because the alternative observables are, as it were, folded
versions of their standard forms, they have a smaller temporal
footprint: that is, they are written as sums of one-way phase
measurements that span a shorter time period. This property can be
advantageous in the presence of instrumental gaps or glitches, which
would then contaminate a smaller portion of the data set; a reduced
temporal footprint means also that a shorter continuous set of phase
data needs to be collected before TDI observables can begin to be
assembled.
\begin{figure}
\includegraphics[width=2.65in,height=3in]{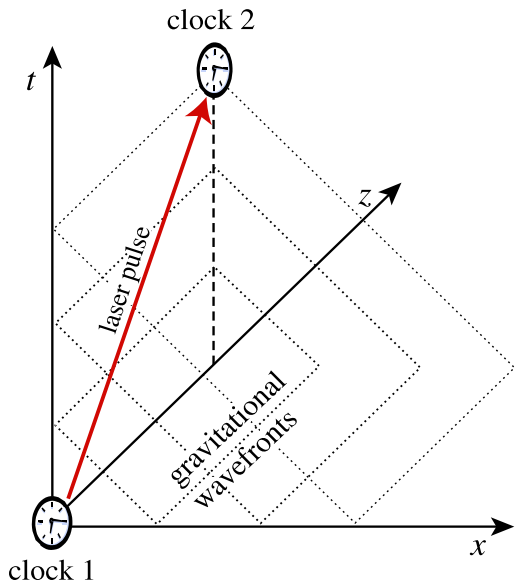}
\caption{In this idealization of the basic time-transport observable
used with LISA, two ideal clocks travel along geodesics, with clock 1
continuously transferring its proper time to clock 2 by way of pulsed
light signals. GWs are measured as the fluctuations in the time of
flight between the clocks (see main text).
\label{fig:gedank}}
\end{figure}

This paper is organized as follows. Section \ref{sec:geometricview}
describes Geometric TDI: in Sec.\ \ref{sec:basictimetransport}, I
introduce the basic GW-sensitive phase measurement; in Sec.\
\ref{sec:tdiprinciple}, I discuss its integration into
laser-noise--canceling observables according to the \emph{Geometric TDI
principle}; in Secs.\ \ref{sec:firstgen} and \ref{sec:secondgen}, I give
a new derivation of the observables of first-generation, modified, and
second-generation TDI, and I interpret them geometrically; in Sec.\
\ref{sec:combenum}, I show how to enumerate exhaustively all possible
observables by representing them as \emph{link strings}; last, in Sec.\
\ref{sec:sixlasers} I extend our formalism, developed for simplicity by
considering only three independent LISA lasers, to the realistic case of
six lasers. Section \ref{sec:survey} reports on the exhaustive survey of
all second-generation TDI observables consisting of up to 24 separate
phase measurements: in Secs.\ \ref{sec:alternative} and
\ref{sec:advantage}, I discuss the alternative forms of the standard
second-generation TDI observables, and present their practical
advantages for the implementation of TDI; in Sec.\ \ref{sec:longer}, I
describe the previously unknown second-generation TDI observables of
length 18 and more. Last, Sec.\ \ref{sec:conclusion} presents my
conclusions. The appendices contain rules and proofs omitted from the
main text, and explicit algebraic expression for the second-generation
TDI observables of length 16.

As customary, I set $G = c = 1$ except where specified otherwise.

\section{A geometric view of time-delay interferometry}
\label{sec:geometricview}

How is LISA an interferometer other than by name? The loosest dictionary
definition of ``interferometer'' (something like ``a device that
combines the signals radiating from a common source, and received at
different locations, or at the same location after traveling different
paths'') does not seem to apply to LISA, whose TDI GW observables are
combinations of the phase-difference measurements between as many as six
laser sources. In fact, interferometry is not needed, strictly speaking,
to measure GWs, but only to remove the otherwise deafening phase noise
produced by the LISA lasers. The basic principle of GW measurement
employed by LISA is noninterferometric, as we can see from the idealized
experimental setup of a \emph{time-transport link} between two ideal
clocks (see Fig.\ \ref{fig:gedank}).

\subsection{The basic time-transport observable}
\label{sec:basictimetransport}

Consider a plane GW propagating across the Minkowski background geometry,
and written in the transverse-traceless gauge \cite{mtw}. The wave is
traveling along the $x$ direction, and has ``$+$'' polarization along
the $y$ and $z$ directions. We can then write the spacetime metric as
$\eta_{\mu \nu} + h^\mathrm{TT}_{\mu \nu}$, where
\begin{equation}
h^\mathrm{TT}_{\mu \nu} = h_+(t + x) [\mathsf{e}_{zz} - \mathsf{e}_{yy}].
\end{equation}
Consider also two ideal clocks 1 and 2, marking their proper times $t_1$
and $t_2$, and sitting at constant spatial coordinates $\vec{p}_1 =
\{0,0,0\}$ and $\vec{p}_2 = \{0,0,L\}$ in the TT frame. In this gauge,
constant-coordinate worldlines are geodesics, so the effect of the GWs
is not to exert forces (as it were) on test particles, but to modulate
the distance between them. By way of light signals, clock 1 is
continuously sending its time $t_1$ to clock 2, where $t_1$ is compared
with the local time $t_2$, yielding the difference
\begin{equation}
\Delta t_{12} = t_2(t) - t_1(t - L_{12}(t)) = L_{12}(t).
\end{equation}
Here $t$ is the TT coordinate time and $L_{12}(t)$ is the time of
flight between the two clocks, as experienced by the laser pulse that
arrives at clock 2 at time $t$. We are assuming that the two clocks have
been synchronized so that in the absence of GWs they both mark the
coordinate time $t$. To first order in $h_+$, $L_{12}(t)$ is
\begin{equation}
L_{12}(t) = L + \frac{1}{2} \int_{t - L}^{t} h_+(t) dt;
\label{eq:timeosc}
\end{equation}
the $x$ coordinate dependence of the GW does not appear in Eq.\
\eqref{eq:timeosc} because the two clocks sit on the same constant-$x$
wavefronts. If the rates of the two clocks remain synchronized ($dt_1/dt =
dt_2/dt$), then the time derivative $d\Delta t_{12}/d t$ is directly
proportional to difference of the GW strains at the events of pulse
reception and emission,
\begin{equation}
\label{eq:timeresponse}
\frac{d \Delta t_{12}}{dt} = \frac{d \Delta t_{12}}{dt_2} =
{\textstyle \frac{1}{2}} [h_+(t) - h_+(t - L)].
\end{equation}
\emph{This is our GW observable.} In the Fourier domain, the (power) response function
of $d \Delta t_{12}/dt$ to GWs is $|1 - \exp(-2\pi i f L)|^2 / 4 =
\sin^2 (\pi f L)$. Thus $d \Delta t_{12}/dt$ is insensitive to GWs of
frequencies $f \ll 1/L$ or $f \simeq k/L$ (with integer $k$). Expression
\eqref{eq:timeresponse} is the basic building block used to derive the
LISA response to GW waves, as well as the Doppler response used in
spacecraft-tracking GW searches \cite{doppler}, and the timing-residual
response used in pulsar-timing searches \cite{pulsar}.

To relate this idealized experimental setup to LISA, we replace the
ideal clocks with the LISA lasers, and obtain proper time by
dividing the lasers' phase by their frequency. Each of the three LISA
spacecraft contains two optical benches oriented facing the other two
spacecraft; on each bench, the appropriately named \emph{phasemeters}
compare the phase of the incoming lasers against the local reference
laser. As written, Eq.\ \eqref{eq:timeresponse} involves a comparison of
laser \emph{frequencies}: we choose to develop our arguments in terms of
these, since it is more convenient to deal with instrumental responses
that are directly proportional to the physical observable of interest
(the GW strain) rather than to its time integral.  Generalizing Eq.\
\eqref{eq:timeresponse} to arbitrary plane-GW and spacecraft geometries,
and adopting a LISA-specific language, we come to the
Estabrook--Wahlquist two-pulse response \cite{estawahl}
\begin{equation}
\label{eq:twopulse}
y_{12}(t) = \frac{1}{2} \frac{\hat{n}^i_{12}(t) \hat{n}^j_{12}(t)
\bigl[h^\mathrm{TT}_{ij}(p^s_2(t),t) -
h^\mathrm{TT}_{ij}
\bigl(p^s_1(t_\mathrm{s}),t_\mathrm{s}\bigr)\bigr]}{1 - \hat{n}^m_{12}(t) k_m}:
\end{equation}
with this indexing, Eq.\ \eqref{eq:twopulse} describes the
frequency-difference measurement performed on spacecraft 2 to compare
the local laser to the laser incoming from spacecraft 1. In this
equation:
\begin{itemize}
\item $k_j$ is the spatial propagation vector of the plane GW;
\item $p_1^s(t)$ and $p_2^s(t)$ are the spatial TT coordinates of the
two spacecraft;
\item $t$ is the time of pulse reception, and therefore of measurement;
\item $t_\mathrm{s}$ is the time of pulse emission, as determined
implicitly by $|p^s_2(t) - p^s_1(t_\mathrm{s})| = t - t_\mathrm{s}$;
\item $\hat{n}^m_{12}(t)$ is the unit vector along the trajectory of the
light pulse (labeled by the time of reception $t$), given by
$\hat{n}^m_{12}(t) = (p^s_2(t) - p^s_1(t_\mathrm{s}))/|p^s_2(t) -
p^s_1(t_\mathrm{s})|$.
\end{itemize}
Equation \eqref{eq:twopulse} is known as the two-pulse response because an
impulsive GW is registered twice in each $y_{ij}(t)$ observable, once
when it impinges on the emitting spacecraft $i$, and once, a time $L$
later, when it impinges on the receiving spacecraft $j$.

In the literature on TDI, it is customary to label the LISA arms by the
index of the \emph{opposite} spacecraft. We shall do so in this paper,
using primed and unprimed indices to denote the \emph{oriented} LISA
arms (with orientation following the direction of laser transmission)
according to the convention\footnote{See the \emph{TDI Rosetta Stone},
first printed in Ref.\ \cite{synthlisa} and updated at
\url{www.vallis.org/tdi}, for a mapping between the different
conventions used in the literature.} $\{1,2,3\} \equiv \{3 \rightarrow
2,1 \rightarrow 3,2 \rightarrow 1\}$ and $\{1',2',3'\} \equiv \{2
\rightarrow 3,3 \rightarrow 1,1 \rightarrow 2\}$. We shall then denote
by $L_l$ the propagation time experienced by a laser pulse traveling
along the oriented arm $l$. We shall also find it useful, at times, to
augment the $y_{ij}$ phase-measurement notation with a middle index,
corresponding to the oriented arm traversed by the laser pulse being
measured. (In fact, the primed or unprimed middle index would be
sufficient to identify the phase measurement completely, and we shall
exploit this property in Sec.\ \ref{sec:combenum} when we represent TDI
observables as \emph{link strings}.) See Fig.\ \ref{fig:tdipaths} for an
example of this convention at work.

\subsection{The Geometric TDI principle}
\label{sec:tdiprinciple}

Unfortunately, GWs cannot be read off directly from the $y_{ij}$
measurements, because the fluctuations $C_i(t)$ of the laser frequencies
(i.e., the laser phase noises) come into the $y_{ij}(t)$ as
\begin{equation}
\label{eq:response}
y_{ij}(t) = C_i(t_\mathrm{s}) - C_j(t) + \mathrm{GWs};
\end{equation}
the LISA lasers have $C_i(t)$ of (single-sided, square-root) spectral
density $\sim 30 \, \mathrm{Hz} / \sqrt{\mathrm{Hz}}$, several orders of
magnitude stronger than the weakest GWs detectable by LISA, which are at
the level of the other two fundamental LISA noises (known together as
\emph{secondary noises}): the shot noise at the phasemeter, as
determined by the power of the lasers and by the distance between the
spacecraft, and the acceleration noise of the proof masses enclosed
within each optical bench, which are used to reference the frequency
measurements to freely falling worldlines. [Equation \eqref{eq:response}
assumes that a single laser is being used on each spacecraft; it is
pedagogical to consider this simplified case first, but we shall
generalize our discussion to the realistic case of six LISA lasers in
Sec.\ \ref{sec:sixlasers}.]
\begin{figure}
\includegraphics[width=3.4in]{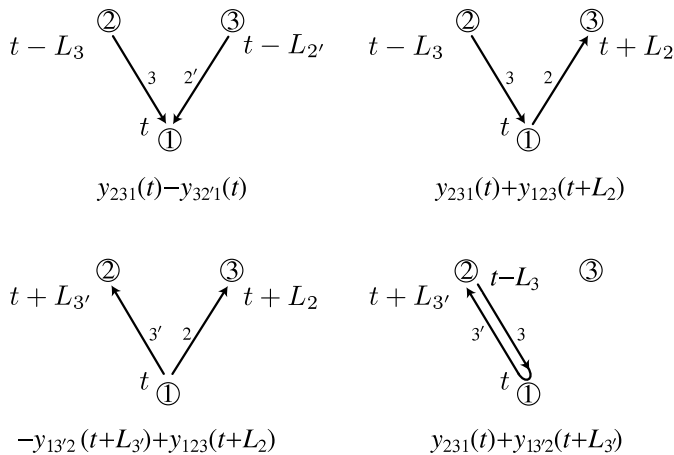}
\caption{These time-delayed sums and differences of two $y_{ij}$
measurements cancel laser phase noise at time $t$. In all of them, two
laser pulses arrive at, or depart from, the same spacecraft at time $t$.
\label{fig:tdipaths}}
\end{figure}

Canceling laser phase noise is where interferometry comes to the rescue.
Look at Fig.\ \ref{fig:tdipaths} for combinations of $y_{ij}$
measurements in which two laser pulses arrive simultaneously at
spacecraft 1 at time $t$, depart simultaneously from spacecraft 1 at
time $t$, or arrive and depart simultaneously to and from spacecraft 1
at time $t$. We subtract the $y_{ij}$ measurements, represented
graphically by arrows, when they share the same event of emission or
reception (i.e., when their arrowtails or arrowheads meet), and we add
them when the receiving spacecraft of one measurement is the emitting
spacecraft of the other (i.e., when arrowtail follows arrowhead).
\emph{In all of these combinations, the laser-frequency noise $C_1(t)$
generated at time $t$ on spacecraft 1 is canceled out} by entering twice
with opposite signs; however, GWs are not canceled (not even at time
$t$), because they come into Eq.\ \eqref{eq:twopulse} with different
$n^m_{ij}$-dependent projection factors. The combinations of Fig.\
\ref{fig:tdipaths} do still contain frequency noise from lasers 2 and 3,
and from times other than $t$; it is however a simple leap to cancel
even those by arranging together more measurements. We then formulate a

\vspace{6pt}
\noindent \textbf{Geometric TDI principle}: to obtain a
laser-noise-canceling GW observable, line up arrows (i.e., $y_{ij}$
measurements) head to head, tail to tail, or head to tail, creating a
closed loop that cancels laser noise at all pulse emission and reception
events. If no arrowhead or arrowtail is left unpaired, the closed loop
represents a linear combination of delayed $y_{ij}$ measurements that
completely cancel the three laser noises $C_i(t)$.
\vspace{6pt}

\begin{figure}
\includegraphics[width=3.4in]{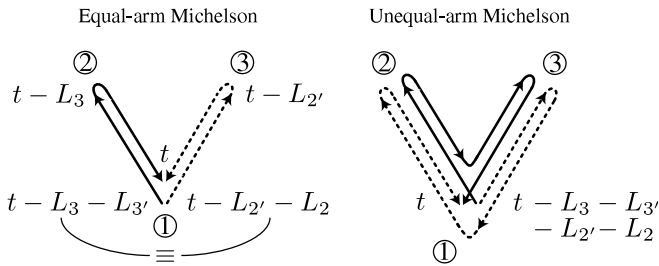}
\caption{\emph{Left.}---The arrows of this closed loop reproduce the
paths of light in a standard equal-arm Michelson interferometer, and the
corresponding time-ordered sum of phase measurements [Eq.\
\eqref{eq:simplemichelson}] reproduces the phase-difference output of
the interferometer. \emph{Right.}---For unequal armlengths,
laser-phase--noise cancellation can be recovered by having both
interfering beams travel along each arm once, building up the same
light-travel time. Compare with Fig.\ of Ref.\ \cite{stea2003}.
\label{fig:michelson}}
\end{figure}

Remarkably, it is usually possible to interpret each closed-loop
combination as the interferometric measurement performed by comparing
the phases of laser beams that follow the paths marked by the arrows.
Let us see an example. The arrows of Fig.\ \ref{fig:michelson} (left
panel) reproduce the paths followed by light in an equal-arm Michelson
interferometer; operating in analogy with Fig.\ \ref{fig:tdipaths}, and
attributing the time $t$ to the final common event of reception at
spacecraft 1, we write the corresponding algebraic expression
\begin{equation}
\label{eq:simplemichelson}
y_{13'2}(t - L_{3}) + y_{231}(t) -
y_{32'1}(t) - y_{123}(t - L_{2'}).
\end{equation}
Here the sum of two time-consecutive $y_{ij}$ observables, such as
$y_{13'2}(t - L_{3})$ and $y_{231}(t)$ [here $L_{3}$ is the light-travel time
between spacecraft 2 and 1], simulates the reflection of the laser off a
mirror: in terms of laser phases, we see that the integral of this sum
reproduces the total phase shift accumulated along the path $1
\rightarrow 2 \rightarrow 1$. By contrast, the head-to-head difference of two such
double arrows simulates a photodetector: it reproduces the difference of
the phase shifts accumulated along the two paths. All in all, Eq.\
\eqref{eq:simplemichelson} shows that the combination of four (one-way)
$y_{ij}$ measurements can \emph{synthesize} the phase-difference output
of a Michelson interferometer, as emphasized by Tinto and Armstrong
\cite{tintoarm}, and shown graphically by Shaddock \cite{stea2003} and Summers \cite{summers}.
Inserting the laser noises $C_i(t)$ in Eq.\ \eqref{eq:simplemichelson}, we get
\begin{align}
[C_1(t - L_{3} - L_{3'}) &- \phantom{[} C_2(t - L_{3})] \nonumber \\
&+ [C_2(t - L_{3}) \phantom{{}_{'}} - C_1(t)] \\
&- [C_3(t - L_{2'}) - C_1(t)] \nonumber \\
- [C_1(t - L_{2'} - L_{2}) &- \phantom{[} C_3(t - L_{2'})] \nonumber,
\end{align}
which sums up to zero in interferometer geometries where\footnote{If the
interferometer is rotating, the Sagnac effect \cite{sagnac} introduces a
distinction between the light-travel times $L_{l}$ and $L_{l'}$ in the
two directions \cite{shaddock2004,cornhell}.} $L_{3} + L_{3'} = L_{2'} + L_{2}$: our equal-arm Michelson combination is then truly laser-noise canceling. It is however sensitive to GWs, as can be seen by inserting Eq.\
\eqref{eq:twopulse} in Eq.\ \eqref{eq:simplemichelson}.

More generally, we can set three simple rules to turn a closed
arrow loop into a combination of $y_{ij}$ measurements that cancels
laser noise:
\begin{enumerate}
\item start at any spacecraft, and write down the appropriate $y_{ij}$
for each arrow, following the loop (going along or against the direction
of each arrow) until all arrows are used up (if more than two heads or
tails meet at any spacecraft, different visiting orders will yield
different observables);
\item use a plus (minus) sign for arrows followed along (against) their
direction;
\item give time arguments to the $y_{ij}$, remembering that measurements
are always made at the receiving spacecraft (at the arrowhead); use the
nominal time $t$ for the first $y_{ij}$, and then add (subtract) the
appropriate $L_l$ for each arrow followed along (against) its
direction.
\end{enumerate}

\subsection{The observables of first-generation TDI}
\label{sec:firstgen}
\begin{figure*}
\includegraphics[width=0.90\textwidth]{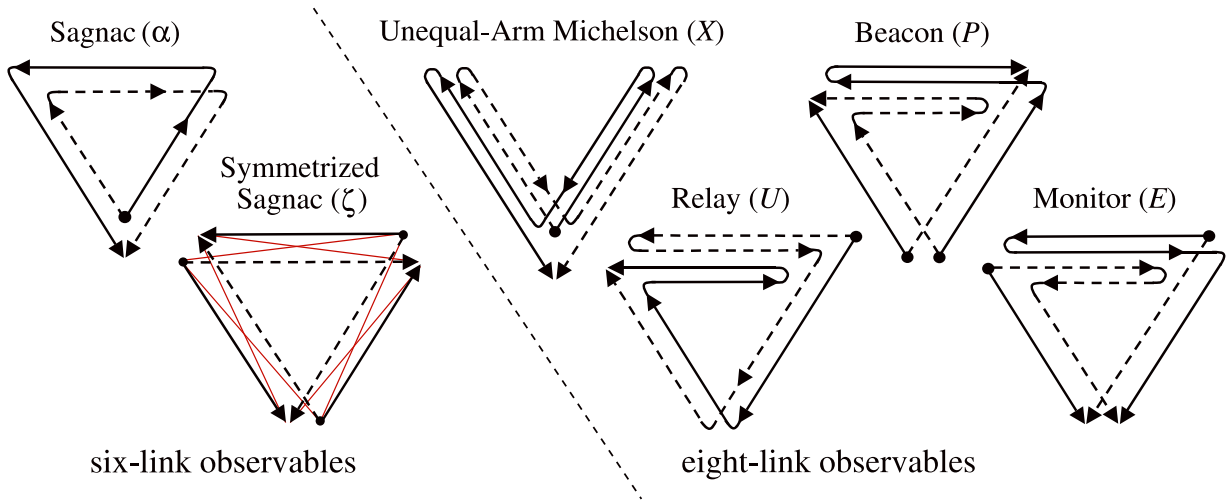}
\caption{Mapping first-generation TDI observables into closed arrow
loops. The Sagnac ($\alpha$), unequal-arm Michelson ($X$), and Relay
($U$) observables of first-generation TDI have a straightforward
interpretation as synthesized two-beam interferometers. More
interestingly, the Beacon ($P$) and Monitor ($E$) observables can be
seen as \emph{four}-beam interferometers, with the four beams combining
into different pairs at the events of initial emission and final
reception. To interpret the symmetrized Sagnac ($\zeta$) observables as
\emph{six}-beam interferometers, three different beam pairings must be
invoked to explain the cancellation of laser noise at emission and
reception, and the relative insensitivity of $\zeta$ to GWs (see main
text).\label{fig:firstgen}}
\end{figure*}

The first laser-noise--canceling combinations for LISA were discovered
using an algebraic (rather than geometric) approach, matching up delayed
$y_{ij}$ measurements in such a way that all laser-noise terms would
cancel. Using this procedure, Tinto, Armstrong, and Estabrook
\cite{tintoarm,firstgen,firstsens} obtained expressions for
\emph{first-generation TDI observables}, which cancel laser noise in
static unequal-arm geometries. These observables are sums of either six
or eight delayed $y_{ij}$ measurements (for short, links). See Fig.\
\ref{fig:firstgen}.

The 6-link observables $\alpha$, $\beta$, $\gamma$ (mapped into each
other by relabeling the spacecraft cyclically) use all six LISA oriented
arms, and measure the phase difference accumulated by two laser beams
traveling around the LISA array in clockwise and counterclockwise
directions: thus, they behave much like a Sagnac interferometer, and are
known as \emph{Sagnac observables}. A related 6-link combination, the
symmetrized Sagnac observable $\zeta$, has the useful property of being
relatively insensitive to GWs in the low-frequency limit.\footnote{Most
interestingly, a GW-insensitive observable allows the observational
distinction of a stochastic GW background from instrumental noise
\cite{zeta}.}

The 8-link observables $X$, $Y$, $Z$ (also mapped into each other by
cyclic spacecraft relabelings) use two of the LISA arms in the two
directions. They are unequal-arm generalizations of the Michelson
observable of Eq.\ \eqref{eq:simplemichelson}: for unequal arms, the
latter would fail to cancel the laser-noise terms from the tails of the
two paths, because $L_3 + L_{3'} \neq L_{2'} + L_{2}$. The solution is
to have both paths go through each arm once (hence the eight terms),
building up the same light-travel time (see the right panel of Fig.\
\ref{fig:michelson}). Related 8-link combinations, known as observables
of the $U$, $P$, and $E$ type, use different sets of four oriented arms
out of six, and have GW sensitivity comparable to the Michelson
combinations \cite{firstgen,firstsens}.

Prior to my work, it was unclear whether the $P$-type and $E$-type
observables could be interpreted as synthesized interferometric
observables.\footnote{The interpretation of $U$ as a synthesized observable was already clear to F.\ B.\ Estabrook (unpublished note).} In Fig.\ \ref{fig:firstgen}, we show that this is possible if we identify \emph{four distinct laser beams}, paired in alternative
ways to cancel laser noise at the path tails (dots) and path heads
(ending arrows). The two path origins are not simultaneous, and neither
are the two path endings.

The symmetrized Sagnac observable $\zeta$, which also defies explanation
as a two-beam synthesized interferometer, can be interpreted as a
\emph{six}-beam interferometer, whereby two different pairings explain
the cancellation of laser noise at emission (dots) and reception
(arrows). Yet another pairing, shown by the thin diagonal lines in Fig.
\ref{fig:firstgen}, explains why $\zeta$ is relatively insensitive to
GWs at low frequencies: in the limit of equal arms, each pair of
parallel arrows represents the difference of two symmetric measurements
$y_{ij}(t)$ and $y_{ji}(t)$ that share the same times of pulse emission
and reception. Taylor-expanding the $h^\mathrm{TT}_{ij}(\ldots;t)$ terms
of Eq.\ \eqref{eq:twopulse} around $t$ and around either $p^l_1$ or
$p^l_2$, we find that $y_{ij}(t) - y_{ji}(t) \propto L^3 h'''_{ij}$. By
contrast, the differences of head-to-tail double arrows that appear in
$X$ sum up to $L^2 h''_{ij}$. Considering monochromatic GWs of frequency
$f_\mathrm{GW}$, we see that the GW response is smaller for $\zeta$ than
for $X$ by a factor\footnote{For unequal interferometer arms, $\zeta
\propto L (\Delta L) h''_{ij}$, so the ratio between the $\zeta$ and $X$
responses is $\sim \Delta L / L \sim 0.01$.} $2 \pi f_\mathrm{GW} L$
($\simeq 0.1$ for $f_\mathrm{GW} = 10^{-3}$ Hz, $\simeq 0.01$ for
$10^{-4}$ Hz). Since the response to the LISA secondary noises is
approximately the same for $\zeta$ and $X$ [as can be seen using Eqs.\
\eqref{eq:transfer}, discussed below], $\zeta$ turns out to be
relatively insensitive to GW.

\subsection{The observables of second-generation TDI}
\label{sec:secondgen}

This interpretation of TDI observables as $2N$-beam synthesized
interferometers is intriguing, but also troubling, since it casts a
suspicion of arbitrariness on the selection of a standard set of
observables, and it complicates exploring the space of all possible
combinations. Fortunately, the application of the tools of modern
algebra to TDI showed that all first-generation observables can be
obtained as algebraic combinations of four generators \cite{dhurandhar}.
This approach was extended \cite{dhurandhar} to \emph{modified} TDI
observables \cite{shaddock2004,cornhell,stea2003,tea2004}, which cancel laser noise in rotating LISA geometries, where the Sagnac effect \cite{sagnac}
introduces a distinction between light-travel times in the two
directions. (The Michelson-, U-, P-, and E-type observables of first-generation TDI are \emph{bona fide} modified TDI observables, if written with the correct primed and unprimed delay indices; by contrast, the Sagnac observables of modified TDI are different, and twice as long as those of first-generation TDI.)

However, the algebraic approach cannot be extended easily to the
observables of \emph{second generation} TDI, which cancel laser noise in
LISA geometries with time-dependent armlengths.\footnote{With
second-generation TDI, the cancellation occurs up to (and including)
terms proportional to $L_l \dot{L}_m$, which is more than sufficient for
realistic LISA spacecraft orbits.} As pointed out by Cornish and
Hellings \cite{cornhell}, in this situation it is necessary to keep
track of the order of retardations: for instance, the unequal-arm
Michelson combination of Fig.\ \ref{fig:firstgen} would translate to
\begin{equation}
\label{eq:modx}
\begin{aligned}
& y_{13'2;322'}(t) + y_{231;22'}(t) + y_{123;2'}(t) + y_{32'1}(t) \\
-\, & y_{231}(t) - y_{13'2;3}(t) - y_{32'1;3'3}(t) - y_{123;2'3'3}(t),
\end{aligned}
\end{equation}
where, using the semicolon notation of Ref.\ \cite{stea2003},
\begin{equation}
\label{eq:semicol}
\begin{aligned}
y_{ilj;r_1}(t) &= y_{ilj}\bigl(t - L_{r_1}\bigr), \\
y_{ilj;r_2r_1}(t) &= y_{ilj}\bigl(t - L_{r_1} - L_{r_2}(t - L_{r_1})\bigr),
\end{aligned}
\end{equation}
and so on: the nominal time $t$ is delayed incrementally starting from
the rightmost delay index $r_1$. (A similar notation with commas instead
of semicolons is used when the armlengths are constant and the order of
the retardations is not important.)

Inserting the laser noises $C_i$ in Eq.\ \eqref{eq:modx}, we see that
they cancel in pairs, except for the terms from the tails of the two
paths,
\begin{equation}
C_{1;3'322'} - C_{1;22'3'3}(t) \simeq
\dot{C}_{1}(t) [t_{;3'322'} - t_{;22'3'3}];
\end{equation}
Taylor-expanding the retardations to first order and keeping only linear
terms in the $\dot{L}_l(t)$, we get
\begin{equation}
\label{eq:modtdires}
\dot{C}_{1}(t) \bigl[
(L_{2'} + L_{2}) (\dot{L}_{3} + \dot{L}_{3'})
-(L_{3} + L_{3'}) (\dot{L}_{2'} + \dot{L}_{2})
\bigr],
\end{equation}
where all the $L_l$ and $\dot{L}_l$ are implicitly evaluated at time
$t$. (More generally, each retardation index $r_i$ generates a residual
term proportional to $L_{r_i} \dot{L}_{r_j}$ for each index $r_{j}$ to
its left.)

In short, much like what happened with the simple Michelson combination
[Eq.\ \eqref{eq:simplemichelson}] for unequal-arm geometries, a
laser-noise residual appears in Eq.\ \eqref{eq:modtdires} because the
light-travel times built up along the two interfering paths are
different; graphically, the tails of the two paths do not match
precisely. This is because, although both paths contain the same set of
links, they do so in different orders, and the retardations
do not commute when the armlengths are time dependent. (This is also the
reason why the algebraic approach becomes arduous for second-generation
TDI, where it involves the solution of polynomial equations for
noncommuting variables.) As in the \emph{upgrade} from equal-arm to
unequal-arm (first-generation) Michelson observables, one solution is to
compose the two paths so that each goes through each arm twice, in
different orders \cite{stea2003}. The residual of the resulting 16-link combination vanishes up to the first Taylor order and to the first
degree in $\dot{L}_l$ (henceforth, \emph{to first order/degree}). This
second-generation unequal-arm Michelson observables (known as $X_1$) may
be written in our notation
\begin{equation}
\label{eq:secx}
\begin{aligned}
& y_{13'2;322'22'3'3}(t) + y_{231;22'22'3'3}(t) \\
+ \, & y_{123;2'22'3'3}(t) + y_{32'1;22'3'3}(t) \\
+ \, & y_{123;2'3'3}(t) + y_{32'1;3'3}(t) + y_{13'2;3}(t) + y_{231}(t) \\
- \, & y_{32'1}(t) - y_{123;2'}(t) - y_{231;22'}(t) - y_{13'2;322'}(t) \\
- \, & y_{231;3'322'}(t) - y_{13'2;33'322'}(t) \\
- \, & y_{32'1;3'33'322'}(t) - y_{123;2'3'33'322'}(t),
\end{aligned}
\end{equation}
which is related to the $X_1$ defined in Refs.\ \cite{stea2003,tea2004} by a
change of sign and by the use of the opposite convention for primed and unprimed
indices.

Second-generation generalizations of all first-generation TDI observables were described by Tinto and colleagues \cite{stea2003,tea2004}. For the analogs of the $\alpha$, $\beta$, $\gamma$, and $\zeta$ observables (which are formally identical to their modified TDI counterparts \cite{shaddock2004,cornhell}, except for the interpretation of the delay indices as noncommuting), laser-noise cancellation is not complete, even to first order/degree: however, the residuals consist of symmetric sums of $L_l \dot{L}_m$ terms that turn out to be small for realistic LISA orbits.

\subsection{The combinatorial enumeration of TDI observables}
\label{sec:combenum}

Our geometric approach to TDI makes it possible to enumerate all the
second-generation TDI observables of given length. The key to this is
the

\vspace{6pt}
\noindent \textbf{Feynman--Wheeler\footnote{``I received a telephone
call one day at the graduate college at Princeton from Professor
Wheeler, in which he said, `Feynman, I know why all electrons have the
same charge and the same mass.' `Why?' `Because, they are all the same
electron!' [\ldots] I did not take the idea that all the  electrons were
the same one from him as seriously as I took the observation that
positrons could simply be represented as electrons going from the future
to the past in a back section of their world lines.'' [R. P. Feynman,
``The Development of the Space-Time View of Quantum Electrodynamics,''
Nobel Lecture, Dec 11, 1965.]} Geometric TDI principle}: any $2N$-beam
Geometric-TDI closed loop can be seen as a \emph{single} beam that
travels forward and backward in time to meet itself back at its origin.
\vspace{6pt}

For instance, the two-beam equal-arm--Michelson combination of Fig.\
\ref{fig:michelson} (left panel), can be interpreted as a single beam
that departs at the initial time $t - L_{3} - L_{3'}$, travels forward
in time to be measured at time $t - L_{3}$, and again travels forward in
time to be measured (and interfere against itself!) at time $t$; the
beam then moves backward in time to be emitted at time $t - L_{2'}$, and
again moves backward in time to be emitted at the original time $t -
L_{2'} - L_{2}$ (equal to $t - L_{3} - L_{3'}$, since the armlengths are
equal). This closes the loop, and cancels laser noise at all junctions
(when we translate graphs to formulas, we must remember to give minus
signs to all the backward-time arrows, drawn dashed in Fig.\
\ref{fig:michelson}).

Once we have established that all $n$-link loops can be represented as a
single loop, we can \emph{enumerate} them combinatorially by choosing a
starting spacecraft and, for $n$ times over, choosing the future or
past time direction, and the leftward (clockwise) or rightward
(counterclockwise) movement direction, in all possible combinations.
Each loop can be denoted by the index of the initial spacecraft,
followed by a string of ``L'' or ``R'' crested by
``$\overrightarrow{}$'' for forward-time arrows and by
``$\overleftarrow{}$'' for backward-time arrows; this notation is
translated easily into strings of link indices crested by their time directions (henceforth, \emph{link strings}). For instance, we would write
$(1)\moverrightarrow{\mathrm{LR}}\moverleftarrow{\mathrm{LR}} \equiv
\moverrightarrow{3'3}\moverleftarrow{2'2}$ and 
$(1)\moverrightarrow{\mathrm{LRRL}}\moverleftarrow{\mathrm{RLLR}}
\equiv
\moverrightarrow{3'322'}\moverleftarrow{33'2'2}$ for the loops in the
left and right panels of Fig.\ \ref{fig:michelson}, respectively.

Not all $3 \times 2^{2n}$ strings with $n$ links correspond to
laser-noise--canceling combinations, because the total light-travel time
accumulated across the loop must be zero (for second-generation TDI,
zero to first order/degree). However, it is quite straightforward to set
simple \emph{closure criteria} that identify the true TDI combinations:

\mylistitem{Pre-TDI interferometry.} For equal-arm geometries, the loop
must end at the initial spacecraft
($\#[\moverrightarrow{\mathrm{L}},\moverleftarrow{\mathrm{R}}] -
\#[\moverrightarrow{\mathrm{R}},\moverleftarrow{\mathrm{L}}] \mod 3 =
0$, where $\#$ denotes the number of occurrences of a symbol in the
string), and it must have a null total light-travel time
($\#[\moverrightarrow{\mathrm{L}},\moverrightarrow{\mathrm{R}}] =
\#[\moverleftarrow{\mathrm{L}},\moverleftarrow{\mathrm{R}}]$). We denote
the combinations that satisfy this property as \emph{closed}.

\mylistitem{First-generation TDI.} For unequal-arm geometries with
generic $L_{l} = L_{l'}$, the loop must end at the initial spacecraft,
and satisfy $\#[\moverrightarrow{l},\moverrightarrow{l'}] =
\#[\moverleftarrow{l},\moverleftarrow{l'}]$ (for $l = 1, 2, 3$), which
yields a null total light-travel time. We denote the combinations that
satisfy this property as $|L|$-\emph{closed}.

\mylistitem{Modified TDI.} For unequal-arm geometries with generic
$L_{l} \neq L_{l'}$, the loop must end at the initial spacecraft and
satisfy $\#[\moverrightarrow{l}] = \#[\moverleftarrow{l}]$ (for $l = 1,
1', 2, 2', 3, 3'$), which yields a null total light-travel time. We
denote the combinations that satisfy this property as $L$-\emph{closed}.

\mylistitem{Second-generation TDI.} For unequal-arm geometries with
generic, time-dependent $L_{l}(t) \neq L_{l'}(t)$, first-order/degree
laser-noise cancellation is obtained for loops that are
$L$-\emph{closed}, and in addition satisfy $\#[\moverrightarrow{l}
\moverrightarrow{\dot{m}},\moverleftarrow{l}\moverleftarrow{\dot{m}}] =
\#[\moverrightarrow{l}\moverleftarrow{\dot{m}},\moverleftarrow{l}
\moverrightarrow{\dot{m}}]$ (with $l, m = 1, 1', 2, 2', 3, 3'$), where a
pair $l \dot{m}$ is counted for each link $\moverrightarrow{l}$
\emph{with itself} and with all the links $\overrightarrow{m}$ and
$\overleftarrow{m}$ to its right, and for each link $\moverleftarrow{l}$
with all the links $\overrightarrow{m}$ and $\overleftarrow{m}$ to its
right. This condition yields a total light-travel time that is null to
first order/degree (see App.\ \ref{app:rule}). For instance, for
$\moverrightarrow{3'3}\moverleftarrow{2'2}$ we count
$\moverrightarrow{3'3'}$,
$\moverrightarrow{3'3}$,
$\moverrightarrow{3'}\moverleftarrow{2'}$,
$\moverrightarrow{3'}\moverleftarrow{2}$,
$\moverrightarrow{33}$,
$\moverrightarrow{3}\moverleftarrow{2'}$,
$\moverrightarrow{3}\moverleftarrow{2}$, and 
$\moverleftarrow{2'2'}$;
hence the counting does not satisfy the property given above. We denote
the combinations that satisfy this property as $\dot{L}$-\emph{closed}.
\lastlistitem

The closure criteria induce useful symmetry properties for the link strings:

\mylistitem{Null bigrams.} The bigrams
$\moverrightarrow{\mathrm{L}}\moverleftarrow{\mathrm{L}}$,
$\moverrightarrow{\mathrm{R}}\moverleftarrow{\mathrm{R}}$,
$\moverleftarrow{\mathrm{L}}\moverrightarrow{\mathrm{L}}$, and
$\moverleftarrow{\mathrm{R}}\moverrightarrow{\mathrm{R}}$ (or
equivalently $\moverrightarrow{l}\moverleftarrow{l}$ and
$\moverleftarrow{l}\moverrightarrow{l}$) always represent combinations
of two $y_{ij}$ measurements that sum up to exactly zero: any $n$-link
string that contains such bigrams represents a combination of length
smaller than $n$.

\mylistitem{Cyclic string shift.} Shifting a string cyclically produces
a combination that has the same closure properties as the original, and
that differs only by an overall time advancement or retardation, and by
an advancement or retardation applied selectively to the shifted terms,
but considered negligible at that closure level (for instance, for an
$L$-closed loop the additional selective retardation would be of leading
order/degree $L_l \dot{L}_m$; for an $\dot{L}$-closed loop, of a higher
order/degree).

\mylistitem{Time and direction reversal.} Simultaneously swapping the
primedness and time direction of all link indices produces a combination
that has the same closure properties as the original, and that differs
from the original by its handedness.

\mylistitem{Cyclic index shift.} Shifting link indices cyclically ($1
\rightarrow 2, 2 \rightarrow 3, 3 \rightarrow 1$) produces a combination
that has the same closure properties as the original, and that differs
only by a relabeling of indices. Noncyclic index permutations, on the
other hand, produce illegal strings (i.e., unconnected loops) unless all
indices change in primedness, in time direction, or in a combination of
the two.\footnote{The reflection-like symmetry considered below Eq.\
\eqref{eq:nonselfref} consists in exchanging two indices and swapping
their primedness, while reversing the time direction of the third
index.}

\mylistitem{String reversal.} Reversing a string while swapping all
``$\overrightarrow{}$'' and ``$\overleftarrow{}$'' symbols (i.e.,
reversing all time directions) produces a combination that has the same
closure properties as the original, and that differs only in sign and by
an overall time advancement or retardation considered negligible at that
closure level.

\mylistitem{Splicing.} Inserting any link string at any
\emph{compatible} point within a link string yields another legal link
string. Here compatibility means that the spacecraft visited by the
first loop at the insertion point (for instance, after
$\moverrightarrow{1}$, spacecraft 2; after $\moverleftarrow{1}$,
spacecraft 3) must be the same as the initial spacecraft of the inserted
loop. The resulting string has \emph{at least} the closure properties
shared by the spliced fragments.
%

For instance, the unequal-arm Michelson ($L$-closed) string
\begin{equation}
\moverrightarrow{{}_1{3'}_2{3}_1{2}_3{2'}}{}_1
\moverleftarrow{{3}_2{3'}_1{2'}_3{2}_1}
\end{equation}
(where the subscripts show the spacecraft visited by the loop before and
after traversing each link) can be spliced at its center with its own
$L$-closed reversal,
\begin{equation}
\label{eq:x1splicing}
\moverrightarrow{{}_1{3'}_2{3}_1{2}_3{2'}}[
\moverrightarrow{{}_1{2}_3{2'}_1{3'}_2{3}}{}_1
\moverleftarrow{{2'}_3{2}_1{3}_2{3'}_1}
]\moverleftarrow{{3}_2{3'}_1{2'}_3{2}_1},
\end{equation}
yielding an $\dot{L}$-closed loop that is in fact the second-generation
TDI Michelson observable.
\lastlistitem

At this point it is also useful to give a rule to translate link strings
into $y_{ij}$ combinations:
\begin{enumerate}
\item Starting at the left end of the string, write a $y_{ij}$
measurement for each index, according to the replacement rules $\{1, 2,
3, 1', 2', 3'\} \equiv \{ y_{32}, y_{13}, y_{21}, y_{23}, y_{31}, y_{12}
\}$; attribute a plus sign for ``$\overrightarrow{}$'' links and a minus
sign for ``$\overleftarrow{}$'' links;
\item while doing this, build the delay sequence to be applied to each
new $y_{ij}$, adding (\emph{from the left}) an advancement index
$\overline{r}$ \emph{before} translating each $\overrightarrow{r}$, and
a retardation index $s$ \emph{after} having translated each
$\overleftarrow{s}$.
\end{enumerate}
For instance, the $\moverrightarrow{3'322'}\moverleftarrow{33'2'2}$
string would translate to
\begin{equation}
\begin{aligned}
& y_{13'2;\overline{3'}}(t)
+ y_{231;\overline{33'}}(t) \\
+ \, & y_{123;\overline{233'}}(t)
+ y_{32'1;\overline{2'233'}}(t) \\
- \, & y_{231;\overline{2'233'}}(t)
- y_{13'2;3\overline{2'233'}}(t) \\
- \, & y_{32'1;3'3\overline{2'233'}}(t)
- y_{123;2'3'3\overline{2'233'}}(t).
\end{aligned}
\label{eq:longadvance}
\end{equation}
To accommodate advancement indices, the semicolon notation of Eq.\
\eqref{eq:semicol} is extended by the advancement rule
\begin{equation}
y_{ilj;\overline{a}}(t) = y_{ilj}(t + \mathit{\Gamma}_{a}),
\end{equation}
with $\mathit{\Gamma}_{a}$ defined as the time experienced by light
propagating along link $a$ for \emph{emission} at time $t$, and given
implicitly in terms of $L_{a}(t)$ by $\mathit{\Gamma}_{a}(t) = L_{a}(t +
\mathit{\Gamma}_{a}(t))$. Retardations and advancements are applied
incrementally starting from the rightmost index.

Equation \eqref{eq:longadvance} appears more complicated than Eq.\
\eqref{eq:modx}, but it encodes essentially the same $y_{ij}$
combination: the two equations are related by a time retardation, as can
be seen by evaluating Eq.\ \eqref{eq:longadvance} at the time
$t_{;3'322'}$. This adds $3'322'$ to the right of the delay sequence for
each $y_{ij}$, and since adjacent pairs $\overline{l}l$ and
$l\overline{l}$ cancel in delay sequences (by the very definition of
$L_l$ and $\mathit{\Gamma}_l$), Eq.\ \eqref{eq:longadvance} then turns
into Eq.\ \eqref{eq:modx}. Conversely, Eq.\ \eqref{eq:secx} may be
obtained by applying the translation rule to the string
$\moverrightarrow{3'322'22'3'3}\moverleftarrow{2'233'33'2'2}$, and
evaluating the resulting expression at the time $t_{;33'2'22'233'}$. A
slightly more complicated version (see App.\ \ref{app:comprule}) of the
rule given above yields $y_{ij}$ combinations that are closer to
standard TDI notation.

\subsection{Extension of Geometric TDI to six-laser LISA configurations}
\label{sec:sixlasers}

The extension of our discussion to LISA configurations with six lasers
introduces three additional laser phase noise variables $C^*_1(t)$,
$C^*_2(t)$, and $C^*_3(t)$, corresponding in Figs.\ \ref{fig:tdipaths}
and \ref{fig:michelson} to the optical benches on the
right-hand side of the spacecraft (if we look toward the center). Equation
\eqref{eq:response} changes accordingly:
\begin{equation}
\begin{aligned}
y_{ij} &= C^*_{i;l} - C_j + \mathrm{GWs}
& \mathrm{for\,unprimed} & \; l(i,j), \\
y_{ij} &= C_{i;l} - C^*_j + \mathrm{GWs}
& \mathrm{for\,primed} & \; l(i,j);
\end{aligned}
\label{eq:ccstar}
\end{equation}
here unprimed and primed link indices $l(i,j)$ correspond to c.c.w.\ and
c.w.\ $y_{ij}$ measurement directions, respectively. As shown in Ref.\
\cite{tea2004}, all arguments and derivations valid with three lasers
can be applied to a six-laser configuration by replacing all $y_{ij}$
with $y^{(6)}_{ij}$ ($\eta_{ij}$ in Ref.\ \cite{tea2004}) defined by
\begin{equation}
\label{eq:sixlaser}
\begin{aligned}
y^{(6)}_{ij} &=
y_{ij} - {\textstyle \frac{1}{2}}(z_{lj;l} - z_{ij;l})
& \mathrm{for\,unprimed} & \; l(i,j), \\
y^{(6)}_{ij} &=
y_{ij} + {\textstyle \frac{1}{2}}(z_{|l|j} - z_{ij})
& \mathrm{for\,primed} & \; l(i,j),
\end{aligned}
\end{equation}
where the $z_{ij}(t)$ are the \emph{back-plane measurements} comparing
the phases of the two lasers on each spacecraft, and where the notation
$|l|$ removes a prime from $l$, if present. Because $z_{ij}(t) = \pm [C^*_j(t) -
C_j(t)]$ [for unprimed and primed $l(i,j)$, respectively], the $C^*_j$
disappear from Eq.\ \eqref{eq:ccstar}, and Eq.\ \eqref{eq:response} is
restored for the $y^{(6)}_{ij}$. This justifies all the developments
reported in this paper also for six-laser LISA configurations.

It should be mentioned in this context that the phase noise from the
random motion of the optical benches enters the $y^{(6)}_{ij}$ with the
same time signature as the laser phase noises, and is therefore also
canceled by TDI. The LISA sensitivity to GWs is then set by the
remaining secondary noises. Adopting the schematization of the
measurement process used in most of the TDI literature, and the notation
used to describe the \emph{Synthetic LISA} simulator \cite{synthlisa},
the response of the $y^{(6)}_{ij}$ to the secondary noises is given by
\begin{equation}
\begin{aligned}
y^{(6)}_{ij} &=
y^\mathrm{op}_{ij} - 2pm_j - pm^*_{i;l} + pm_{i;l}
& \mathrm{for\,unprimed} & \; l(i,j), \\
y^{(6)}_{ij} &=
y^\mathrm{op}_{ij} - pm^*_j - pm_j
& \mathrm{for\,primed} & \; l(i,j),
\end{aligned}
\label{eq:transfer}
\end{equation}
where $y^\mathrm{op}_{ij}$ is the optical-path noise in the $y_{ij}$
phase measurement, and $pm_i$ and $pm^*_i$ are the velocity noises of
the two proof masses aboard spacecraft $i$. Because $pm_{i;l} - pm_i$
and $pm^*_{i;l} - pm^*_i$ have the same time signature as laser phase
noises, they are canceled in TDI observables; thus, all retardations can
be removed from the unprimed-$l(i,j)$ expression of $y^{(6)}_{ij}$,
casting it to the same form as its primed-$l(i,j)$ counterpart.

\section{A geometric survey of second-generation TDI observables}
\label{sec:survey}

I have written a computer program to list all the second-generation TDI
observables consisting of 24 or fewer $y_{ij}$ measurements. For each
even length $n$, this was achieved by enumerating all $2^{2n}$ possible
$\mathrm{LR}$ strings, and checking each of them for $\dot{L}$-closure,
according to the counting rule given in Sec.\ \ref{sec:combenum}.
Already for 24-link strings, the combinatorial space is huge, and an
exhaustive search required more than 10,000 CPU hours. The resulting
list of observables was then reduced to a minimal set by removing all
the quasi-duplicates that differ only by a sign or by a cyclic string
shift. I have kept as distinct the observables that differ by a cyclic
index shift (in first-generation TDI, this would correspond to counting
$X$, $Y$, and $Z$ as separate observables). The reduced list of TDI
observables is available at the webpage \url{www.vallis.org/tdi},
annotated with their temporal footprint (see Sec.\ \ref{sec:advantage}),
number of beams, type, and splicing composition (see Secs.
\ref{sec:alternative} and \ref{sec:longer}).
\squeezetable
\begin{table*}
\begin{tabular}{r||r|r||r|r|r|r||r|r|r|r|r}
links & comb.\ space & unique obs.\ &
$X$-type & $P,E$-type & $U$-type & other &
2-beam & 4-beam & 6-beam & 8-beam & 10-beam and higher \\
\hline
16 & $4 \times 10^{9\phantom{1}}$ &    48 &  12 & 18 & 18 &     0 &   3 &   27 &    0 &   18 & 0 \\
18 & $7 \times 10^{10}$           &   192 &   0 & 12 & 12 &   168 &   6 &   24 &  102 &   60 & 0 \\
20 & $1 \times 10^{12}$           &   660 &  24 & 18 & 18 &   600 &  12 &  114 &  276 &   90 & 168 \\
22 & $2 \times 10^{13}$           &  2412 &   0 & 36 & 36 &  2340 &  30 &  264 &  732 &  792 & 594 \\
24 & $3 \times 10^{14}$           & 12585 & 144 & 90 & 90 & 12261 &  99 &  945 & 2676 & 4566 & 4299	
\end{tabular}
\caption{Geometric survey of second-generation TDI observables.
Quasi-duplicates that differ only by a sign or by a time advancement are
not counted. The $X$, $U$, $E$, and $P$ types represent generalizations
of the 8-link observables of the same name: $X$-type (Michelson)
observables use two arms in both directions, $U$-type observables use
four oriented arms in a \emph{relay} configuration, $E$-type and
$P$-type use four oriented arms in \emph{beacon} and \emph{monitor}
configurations; the observables reported under \emph{other} use either
five or six oriented arms. The number of beams corresponds to the number
of contiguous substrings with the same time direction, minimized with
respect to cyclic string shifts. All the observables tallied in this
table are available \emph{in extenso} at the webpage
\protect\url{www.vallis.org/tdi}. \label{tab:second}}
\end{table*}

My results are tallied in Table \ref{tab:second}. Here the $X$, $U$,
$E$, and $P$ types represent generalizations of the 8-link observables
of the same name: $X$-type (Michelson) observables use two arms in both
directions, $U$-type observables use four oriented arms in a
\emph{relay} configuration, $E$-type and $P$-type use four oriented arms
in \emph{beacon} and \emph{monitor} configurations; the observables
tallied under \emph{other} use either five or six oriented arms. Here
are the highlights of the survey, which are discussed in more detail in
the following sections.
\begin{itemize}
\item I find that the \emph{shortest} second-generation TDI observable
has length 16. By contrast, modified-TDI observables begin at length 8.
\item I recover all\footnote{In fact, our $U$-type combinations do not
include the $U_1$ observable given in Ref.\ \cite{tea2004}, which
achieves laser-noise cancellation at the approximate time $t - 4L$
through the sum of \emph{four} distinct $y_{ij}$ measurements; by
contrast, in Geometric TDI laser noise is always canceled \emph{by
construction} between \emph{pairs} of phase measurements. However, the
$U_1$ of Ref.\ \cite{tea2004} has almost the same temporal structure as
our
$\moverrightarrow{1'12'3'}\moverleftarrow{1}\moverrightarrow{2'3'1'1}
\moverleftarrow{3'2'1'11'3'2'}$ (keep in mind that the primedness of our
indices is the opposite of Ref.\ \cite{tea2004}).\label{notetea}} the
\emph{known} 16-link second-generation TDI observables, previously
obtained by Tinto and colleagues \cite{stea2003,tea2004} by applying
commutator-like delay operators to the 8-link observables of modified
TDI. From a geometric viewpoint, all the 16-link observables can be
understood as \emph{self-splicings} of the 8-link observables of the
same type. This shows that the former reduce to finite time differences
of the latter, up to time shifts of the first order/degree. It follows
that the second-generation TDI observables have the same sensitivity of
the modified TDI observables of the same type,\footnote{Neglecting of
course the fact that the modified-TDI observables would not cancel laser
noise in a flexing LISA.} not only in the equal-arm limit, but
unconditionally.

See Sec.\ \ref{sec:alternative} for details.
\item In addition, I obtain \emph{alternative forms} of the known
16-link observables. The alternative forms use a larger number of beams
(e.g., four beams for $X$, as opposed to the standard two), or a
different allocation of links in the beams (e.g., $5+4+3+4$ or $6+4+2+4$
for $U$, as opposed to the standard $7+4+1+4$ structure). The
alternative forms, too, can be understood as \emph{self-splicings} of
the 8-link modified TDI observables of the same kind.

The alternative forms have the same sensitivity to GW signals as the
original forms in idealized measurement conditions, but they can improve
on them when realistic aspects (such as quantization of the phasemeter output and technical
noises) are taken into account. In addition, the alternative forms have
a reduced temporal footprint (the difference between the times of the
earliest and latest phase measurements involved in their construction);
this feature can be advantageous in the presence of gaps or glitches in
the $y_{ij}$ data, because it reduces the extent of defect propagation
to the TDI time series.

See Secs.\ \ref{sec:alternative} and \ref{sec:advantage} for details.
Appendix \ref{app:secondsixteen} gives explicit algebraic expressions for all the 16-link observables in terms of the $y_{ij}$ measurements.
\item Second-generation TDI observables are found in \emph{increasing
numbers} at lengths 18, 20, 22, and 24. A minority are of the $X$, $U$,
$E$, or $P$ types, while most use either five or six oriented arms.

All 18-to-24--long observables can be understood as splicings of
modified-TDI observables of length 8 to 18, sometimes with the inclusion
of null bigrams; most, but not all, are self-splicings. I conjecture
that all second-generation TDI observables of any length can be
generated as splicings of two modified-TDI observables.

See Sec.\ \ref{sec:longer} for details.
\item Up to length 24, I do not find any $\dot{L}$-closed observables of
the $\zeta$ type (defined as having suppressed, but nonzero, GW
response at low frequencies). I conjecture that the $\zeta$ type is
incompatible with $\dot{L}$-closure. This does not exclude the existence
of non-$\dot{L}$-closed $\zeta$-type observables (such as the $\zeta_1$,
$\zeta_2$, and $\zeta_3$ described by Tinto and colleagues
\cite{tea2004}) that do not cancel laser noise to first order/degree,
but bring it sufficiently below the LISA secondary noise to be useful in
practice.
\end{itemize}

\subsection{Structure and sensitivity of the 16-link observables}
\label{sec:alternative}

The standard second-generation TDI $X$ observable is
\begin{equation}
X^{16,2}_1:
\moverrightarrow{3'322'22'3'3}
\moverleftarrow{2'233'33'2'2}
\;\;\; (\mathrm{2\;beams}),
\end{equation}
which is related to the $X_1$ defined in Eq.\ \eqref{eq:secx} by
$X^{16,2}_1 \equiv X_{1;\overline{33'2'22'233'}}$. The alternative forms
found in our geometric survey are
\begin{equation}
\begin{aligned}
X^{16,4,+1}_1: & \,
\moverrightarrow{3'322'}
\moverleftarrow{33'2'233'}
\moverrightarrow{22'3'3}
\moverleftarrow{2'2} \\
X^{16,4,-1}_1: & \,
\moverrightarrow{3'3}
\moverleftarrow{2'233'}
\moverrightarrow{22'3'322'}
\moverleftarrow{33'2'2}
\end{aligned}
\;\;\; (\mathrm{4\;beams}),
\label{eq:alternativex}
\end{equation}
which differ between themselves only by handedness, and
\begin{equation}
X^{16,4,0}_1: \moverrightarrow{22'22'}
\moverleftarrow{33'2'2}
\moverrightarrow{3'33'3}
\moverleftarrow{2'233'}
 \;\;\; (\mathrm{4\;beams}),
\end{equation}
which turns out to have vanishing response in the equal-arm limit to
both noise and GWs, at all frequencies. To see that $X_1$ and the
$X^{16,4,\pm 1}_1$ have all the same GW sensitivity as the 8-link
modified-TDI $X$ (neglecting of course the fact that $X$ would not
cancel laser noise in a flexing LISA), we reason as follows.

As we have learned in Eq.\ \eqref{eq:x1splicing}, $X_1$ can be
interpreted as a self-splicing of $X$ with its reversal. If we take $X$
to be defined by Eq.\ \eqref{eq:modx}, we see that
\begin{equation}
\begin{aligned}
\moverrightarrow{3'322'}\moverleftarrow{33'2'2}
&\equiv X_{;\overline{2'233'}}, \\
\moverrightarrow{22'3'3}\moverleftarrow{2'233'}
&\equiv -X_{;\overline{33'2'2}}.
\end{aligned}
\end{equation}
Since the time at the splicing point in
$\moverrightarrow{3'322'}\moverleftarrow{33'2'2}$ is
$t_{;\overline{2'233'}}$, from Eq.\ \eqref{eq:x1splicing} we see that
\begin{equation}
\label{eq:torewrite}
\moverrightarrow{3'322'22'3'3}\moverleftarrow{2'233'33'2'2}
\simeq X_{;\overline{2'233'}} - X_{;\overline{33'2'22'233'}};
\end{equation}
here the symbol ``$\simeq$'' denotes equality up to selective delays or
advancements of order $L_l \dot{L}_m$ in the $y_{ij}$, not specified by
the formal delay strings. [In this case, these spurious delays appear
because $\moverrightarrow{22'3'3}\moverleftarrow{2'233'}$ is only
$L$-closed, so the time at the beginning and at the end of the inserted
string is different by terms of order $L_l \dot{L}_m$; consequently, the
last four $y_{ij}$ observables of
$\moverrightarrow{3'322'}\moverleftarrow{33'2'2}$ are really evaluated
at the time $t_{;3'322'\overline{33'2'22'233'}}$, not just
$t_{;\overline{2'233'}}$.]

Rewriting Eq.\ \eqref{eq:torewrite} in terms of $X_1$ and reabsorbing
the time advancements by evaluating the equation at time
$t_{;3'322'22'3'3}$, we find
\begin{equation}
\label{eq:firsttosecond}
X_1 \simeq X_{;22'3'3} - X.
\end{equation}
Thus, up to delays of first order/degree, self-splicings produce finite
differences of observables. (Indeed, it is a well-known fact in the
literature on second-generation TDI observables that the standard
16-link observables are approximately equal to finite differences of the
standard 8-link observables of the same type.)

Now, because the individual $y_{ij}$ measurements respond
linearly\footnote{Linearity is always assumed in the model of
measurement used to derive TDI. Significant nonlinearity in the phase
measurements has been explored little, but would probably prove very
detrimental to the delicate cancellation of laser phase noise achieved
by TDI.} to GWs and to all instrumental noise sources, the strain
sensitivity of $X$ to monochromatic sources of frequency $f$ at a given
sky position is proportional to $\tilde{X}^n(f) /
\tilde{X}^\mathrm{GWs}(f)$, where $\tilde{X}^n(f)$ is the
(square-root) spectral density of noise in $X$, and
$\tilde{X}^\mathrm{GWs}$ is the Fourier transform of the GW response
function. The constant of proportionality is $\mathrm{SNR} /
\sqrt{T_\mathrm{obs}}$, where SNR is the fiducial signal-to-noise ratio
at which the sensitivity is defined, and $\sqrt{T_\mathrm{obs}}$ is the
time duration of the observation. Combining the Fourier-transform
time-shifting property with Eq.\ \eqref{eq:firsttosecond}, and
considering that first order/degree terms can be neglected for secondary
noises and GWs (which are much weaker than the laser phase noises), we
see that $X_1$ must have the same sensitivity as $X$:
\begin{equation}
\frac{
\tilde{X}_1^n(f)
}{
\tilde{X}_1^\mathrm{GWs}(f)
}
=
\frac{
(e^{2\pi i f \Delta t} - 1) \tilde{X}^n(f)
}{
(e^{2\pi i f \Delta t} - 1) \tilde{X}^\mathrm{GWs}(f)
} \simeq
\frac{
\tilde{X}^n(f)
}{
\tilde{X}^\mathrm{GWs}(f)
},
\label{eq:sensitive}
\end{equation}
with $\Delta t = t_{;22'3'3} - t$. This is true generically for any
spacecraft geometry, and not just in the equal-arm limit.

To see that $X^{16,4,+1}_1$ and $X^{16,4,-1}_1$, too, have the same sensitivity to
GWs as $X$ (in the limit of perfect laser phase noise cancellation), it
is then sufficient to show that they are
self-splicings of $X$ and of its (cyclically shifted) reversal:
\begin{equation}
\begin{aligned}
X^{16,4,+1}_1: & \;
\moverrightarrow{3'322'[33'}
\moverleftarrow{2'233'}|
\moverrightarrow{22'}]
\moverleftarrow{3'32'2}, \\
X^{16,4,-1}_1: & \;
\moverrightarrow{3'3}[
\moverleftarrow{2'233'}|
\moverrightarrow{22'3'3]22'}
\moverleftarrow{33'2'2}.
\end{aligned}
\end{equation}

Moving on to the $U$ type, we see (for instance) that the three
second-generation $U$-type observables that use the oriented arms $1$,
$1'$, $2$, and $3$,
\begin{equation}
\label{eq:utypelong}
\begin{aligned}
& \moverrightarrow{3211'}\moverleftarrow{231[23}|
\moverrightarrow{1'132}\moverleftarrow{1'1]1'}, \\
& \moverrightarrow{3211'}\moverleftarrow{23}
[\moverrightarrow{1'132}\moverleftarrow{1'123]11'}, \\
& \moverrightarrow{3211'[132}|\moverleftarrow{1'123}
\moverrightarrow{1'}]\moverleftarrow{2311'},
\end{aligned}
\end{equation}
are generated by the self-splicings of the modified TDI $U$-type
observable that uses the same oriented arms,
$\moverrightarrow{3211'}\moverleftarrow{2311'}$. Thus, the modified TDI
and second-generation TDI $U$-type observables have the same GW
sensitivity (in the limit of perfect laser phase noise cancellation).
Similar arguments hold for the $P$- and $E$-type observables.

Altogether, we find empirically that all the 16-link second-generation TDI
observables can be generated as self-splicings of the 8-link modified
TDI observables of the same kind. Conversely, it can be proved that all
self-splicings of $L$-closed observables are $\dot{L}$-closed (see App.\
\ref{app:closedproof}).

\subsection{Advantages of the alternative 16-link observables}
\label{sec:advantage}
\begin{figure}
\includegraphics[width=3.3in]{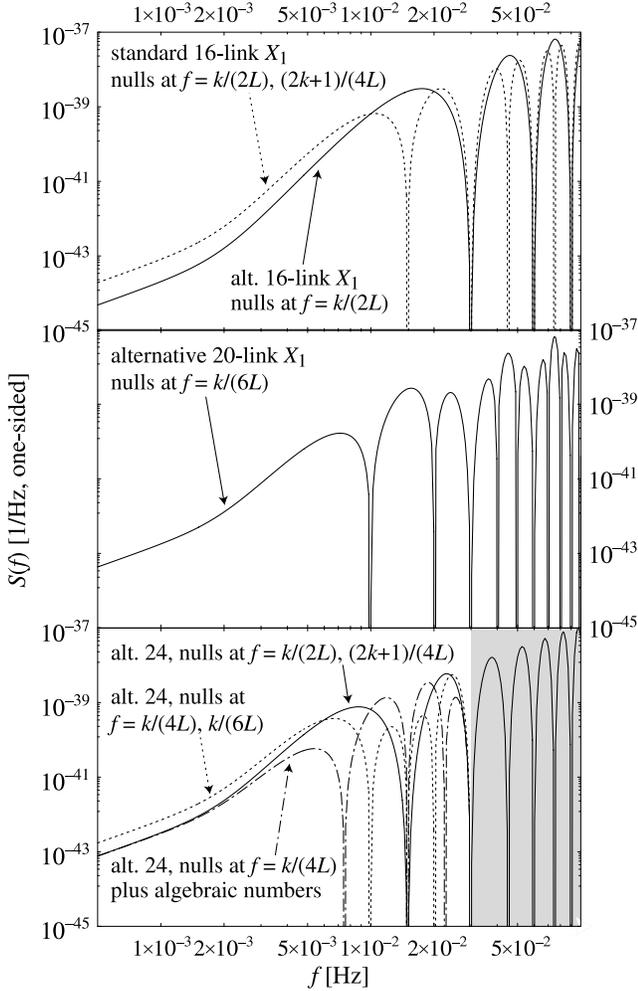}
\caption{TDI response of the second-generation $X$-type variables to the
fundamental secondary noises, according to Eq.\ \eqref{eq:transfer},
assuming equal arms and proof-mass optical-path noise spectral densities
given by $S^\mathrm{pm}_i = 2.5 \times {10^{-48}} [f/\mathrm{Hz}]^{-2}$
Hz$^{-1}$ and $S^\mathrm{op}_i = 1.8 \times {10^{-37}}$ [f/Hz]$^2$
Hz$^{-1}$ (following Ref.\ \cite{synthlisa}). The top panel shows the
noise response of the standard 16-link $X_1$ observable, as compared to
the response of the alternative 16-link forms of Eq.\
\eqref{eq:alternativex}, which have half as many nulls. The middle panel
shows the noise response of the two-beam, 20-link $X_1$ observables of
Eq.\ \eqref{eq:twobeam20}. The bottom panel shows the three new noise
responses found for 24-link $X_1$ observables (solid and dashed: self-
and nonself-splicings of 12-link observables; dash-dotted:
self-splicings of 8-link observables, with inclusions). To avoid visual
clutter, two of the 24-link $X_1$ curves are not plotted beyond $f =
1/(2L)$ (shaded region).\label{fig:sens}}
\end{figure}

As mentioned above, the alternative forms of the 16-link observables can
have a \emph{smaller time footprint} than the standard forms. For
instance, the standard $X_1(t)$ of Eq.\ \eqref{eq:secx} involves 16
$y_{ij}$ measurements taken\footnote{Although the nominal times of the
$y_{ij}$ are displaced only by $\simeq 7L$, we must remember that in a
six-laser LISA configuration some of the $y^{(6)}_{ij}$ contain one
additional delay, as given by Eq.\ \eqref{eq:sixlaser}.} within the
interval $[\mathrm{min}(t_{;3'322'22'3'3},t_{;22'3'33'322'}),t]$, for a
time span $\simeq 8L$, and each single $y_{ij}$ measurement appears in
$X_1$ at times displaced by as much as $\simeq 6L$. Thus, $X_1$ will be
unavailable during the first and the last $\simeq 136$ s (i.e., $8L$)
within each LISA data-taking period. Moreover, a data gap in a single
$y_{ij}$ measurements will appear in the $X_1$ time series at four
distinct times spanning $\simeq 6L$. By contrast, the alternative forms
$X_1^{16,4,\pm 1}$ involve 16 $y_{ij}$ measurements taken within a time
interval of span $\simeq 6L$, and the single $y_{ij}$ appear at times
displaced by at most $\simeq 4L$. The gain is significant, if not
dramatic.

The alternative forms for the $U$-, $E$- and $P$-type variables also
yield footprints gains with respect to their standard forms (from
$\simeq 7L$ to $\simeq 5L$ for $U$, and from $\simeq 5L$ to $\simeq 4L$
for $E$ and $P$). These gains are possible because the alternative
observables are obtained, loosely speaking, by \emph{folding} the
standard versions in time, using both time advancements and retardations,
as opposed to retardations only, to arrange the $y_{ij}$ measurements
so that laser phase noise is canceled at all emission and reception
events.

Another advantage of the alternative forms is that they can yield an
improvement in GW sensitivity in realistic measurement conditions. In
the top panel of Fig.\ \ref{fig:sens}, the dashed curve shows the power
spectral density (PSD) of secondary noise for the standard $X_1$
observable, drawn in the limit of equal armlengths. Following Ref.\
\cite{synthlisa}, we assume that secondary noise consists entirely of
proof-mass noise (idealized as stationary and Gaussian, with PSD
$S^\mathrm{pm}_i = 2.5 \times {10^{-48}} [f/\mathrm{Hz}]^{-2}$
Hz$^{-1}$) and of optical-path noise (also stationary and Gaussian, with
PSD $S^\mathrm{op}_i = 1.8 \times {10^{-37}}$ [f/Hz]$^2$ Hz$^{-1}$). As
discussed around Eq.\ \eqref{eq:sensitive}, the sensitivity to GWs
(averaged over noise realizations) is computed by dividing the
secondary-noise rms power by the GW transfer function.\footnote{Roughly
speaking, the GW transfer function for the standard $X_1$ is found by
inserting Eq.\ \eqref{eq:twopulse} into Eq.\ \eqref{eq:secx} and
Fourier-transforming, assuming a monochromatic source at a fixed sky
location. The resulting transfer function is usually integrated over sky
locations.} See, for instance, Ref.\ \cite{firstsens} for plots of the
GW sensitivities of the first-generation $X$-, $U$-, $P$-, and $E$-type
observables, common also (as discussed above) to the second-generation
observables of the same type.

Although the noise PSD has nulls at multiples of the inverse armlength
light-travel time [for the standard $X_1$, the nulls are at $f = k/(4
L)$], the sensitivity to GWs remains finite in idealized conditions,
because the GW transfer function displays zeros of the same order at the
same frequencies. In reality, we should expect a degradation of
sensitivity at these frequencies, because noise as a whole can only drop
to the level of uncanceled laser noise,\footnote{However, the
simulations of Ref.\ \cite{tdir} seem to suggest that uncanceled laser
noise, under the assumption of perfectly linear responses, will display
the same nulls as the fundamental secondary noises.} or of other
technical noises (such as quantization noise), effectively filling in
the nulls. The nulls are unwelcome also because they imply that a very
large dynamic range is needed for sensitive measurement at those
frequencies. The alternative forms $X_1^{16,4,\pm 1}$ improve on this
situation, because their noise PSD and GW transfer functions\footnote{In
the equal-arm limit, $S_{n}^\mathrm{orig,16} = 4 \cos^2 (2 \pi f L)
S_{n}^\mathrm{alt,16}$.} have half as many nulls [at $f = k/(2 L)$], as
shown by the solid curve in the top panel of Fig.\ \ref{fig:sens}.

Similar gains are found for the other 16-link second-generation TDI
observables. The standard $U$-type observable of Ref.\ \cite{tea2004},
which we may represent as\footnote{As discussed in note \ref{notetea},
the observables are not identical: however, they do have the same
secondary noise PSD and GW transfer function.}
$\moverrightarrow{3211'132}\moverleftarrow{1'123}\moverrightarrow{1'}
\moverleftarrow{2311'}$, has nulls at $f = k/(3L)$, while the alternative
form
$\moverrightarrow{3211'}\moverleftarrow{23123}\moverrightarrow{1'132}
\moverleftarrow{1'11'}$ has nulls only at $f = k/L$. The distribution of
the nulls improves also for all the alternative forms of the $E$- and
$P$-type observables.

\subsection{Longer observables} \label{sec:longer}

Although the size of the combinatorial space of LR strings scales as
$2^{2n}$ with increasing string length $n$, we find empirically that
the number of valid
$\dot{L}$-closed combinations grows only as $2^n$ (at least up to $n =
24$); by contrast, the number of the less constrained $L$-closed
combinations grows roughly as $2^{3n/2}$ (at least up to $n = 16$).

The majority of the longer second-generation observables use either five
or six LISA oriented arms, and therefore do not belong to any of the
$X$, $U$, $E$, or $P$ types. We do however find new forms for these.
Throughout this section, we shall consider examples of the $X$ type:
findings and conclusions are similar for the other types.

At length 20, we find two $X$-type observables with two beams,
\begin{equation}
\begin{aligned}
& \moverrightarrow{3'322'\mathbf{22'}[22'3'3}\moverleftarrow{2'233']
\mathbf{2'2}33'2'2}, \\
& \moverrightarrow{3'322'\mathbf{3'3}[22'3'3}\moverleftarrow{2'233']
\mathbf{33'}33'2'2},
\end{aligned}
\label{eq:twobeam20}
\end{equation}
which can be interpreted as the self-splicing of the 8-link modified TDI
observable $\moverrightarrow{3'322'}\moverleftarrow{33'2'2}$ with its
reversal $\moverrightarrow{22'3'3}\moverleftarrow{2'233'}$, after the
\emph{insertion of the double null bigrams}
$\moverrightarrow{22'}\moverleftarrow{2'2}$ and
$\moverrightarrow{3'3}\moverleftarrow{33'}$ [shown in bold in Eq.\
\eqref{eq:twobeam20}]. Because the $y_{ij}$ measurements corresponding
to the null bigrams sum up to zero to first order/degree, we can repeat
the arguments of Sec.\ \ref{sec:alternative} to see that the 20-link
$X_1$ observables of Eq.\ \eqref{eq:twobeam20} have the same sensitivity
as the standard 16-link form. Moreover, reasoning along the lines of
App.\ \ref{app:closedproof}, we can prove that the self-splicings of
$L$-closed strings are $\dot{L}$-closed even with the inclusion of null
bigrams.

At length 20 we find also alternative $X_1$ observables with four beams,
\begin{equation}
\begin{aligned}
& \moverrightarrow{3'322'\mathbf{22'}}
[\moverleftarrow{33'}|\moverrightarrow{22'3'3}
\moverleftarrow{2'2]\mathbf{2'2}33'2'2}, \\
& \moverrightarrow{3'322'\mathbf{3'3}
[3'3}\moverleftarrow{2'233'}|\moverrightarrow{22'}]
\moverleftarrow{\mathbf{33'}33'2'2}, \\
& \moverrightarrow{3'3\mathbf{3'3}
[22'3'3}\moverleftarrow{2'233']\mathbf{33'}}
\moverrightarrow{22'}\moverleftarrow{33'2'2},  \\
& \moverrightarrow{3'322'}\moverleftarrow{33'}
\moverrightarrow{\mathbf{22'}[22'3'3}\moverleftarrow{2'233']\mathbf{2'2}2'2},
\end{aligned}
\end{equation}
and with six beams,
\begin{equation}
\begin{aligned}
& \moverrightarrow{3'322'}\moverleftarrow{33'}
\moverrightarrow{\mathbf{22'}}[\moverleftarrow{3'3}
|\moverrightarrow{22'3'3}\moverleftarrow{2'2]\mathbf{2'2}2'2}, \\ 
& \moverrightarrow{3'322'}\moverleftarrow{33'\mathbf{33'}
[33'}|\moverrightarrow{22'3'3}\moverleftarrow{2'2}]
\moverrightarrow{\mathbf{3'3}}\moverleftarrow{2'2},
\end{aligned}
\end{equation}
all of which are again self-splicings of
$\moverrightarrow{3'322'}\moverleftarrow{33'2'2}$ with two null-bigram
inclusions. The secondary-noise PSD of the two-beam, 20-link $X_1$
observables of Eq.\ \eqref{eq:twobeam20} is shown in the middle panel of
Fig.\ \ref{fig:sens}, and has nulls at $f = k/(6L)$. This PSD is not
found for any of the 16-link $X_1$ observables. The four-beam
observables have the secondary-noise PSD of the standard 16-link $X_1$
(the dashed curve in the top panel of Fig.\ \ref{fig:sens}), while the
six-beam observables have the PSD of the alternative 16-link $X_1$ (the
solid curve in the top panel of Fig. \ref{fig:sens}).

We find more $X$-type observables at length 24. Some of these, such as
\begin{equation}
\moverrightarrow{3'322'\mathbf{22'22'}[22'3'3}
\moverleftarrow{2'233']\mathbf{2'22'2}33'2'2},
\end{equation}
are self-splicings of the 8-link modified TDI $X$ with quadruple
null-bigram inclusions, and have the same GW sensitivity as the standard
16-link $X_1$. They can have the same secondary-noise PSDs as the 16-
and 20-link $X_1$ observables, or a new PSD (the dash-dotted curve in
the bottom panel of Fig.\ \ref{fig:sens}) with nulls at $f = k/(4L)$ and
at frequencies given by third-degree algebraic numbers.

Other 24-link $X$-type observables, such as
\begin{equation}
\begin{aligned}
& \moverrightarrow{3'322'22'[22'22'3'3}\moverleftarrow{2'22'233']33'2'22'2}, \\
& \moverrightarrow{3'33'322'[22'3'33'3}\moverleftarrow{2'233'33']33'33'2'2},
\label{eq:self12}
\end{aligned}
\end{equation}
are self-splicings of the \emph{12-link} modified TDI $X$-type
observables $\moverrightarrow{3'322'22'}\moverleftarrow{33'2'22'2}$ and
$\moverrightarrow{3'33'322'}\moverleftarrow{33'33'2'2}$. The 24-link
observables of Eq.\ \eqref{eq:self12} can be shown to have the same GW sensitivity as the standard
16-link $X_1$ \emph{in the equal-armlength limit}. To see this, we parse
the 8- and 12-link $X$-type variables as self-splicings of the pre-TDI
(simply closed) observable $\moverrightarrow{3'3}\moverleftarrow{2'2}$,
respectively without and with double--null-bigram inclusions:
\begin{equation}
\begin{gathered}
\moverrightarrow{3'3[22'}\moverleftarrow{33']2'2}, \\
\moverrightarrow{3'3\mathbf{22'}[22'}\moverleftarrow{33']\mathbf{2'2}2'2}, \\
\moverrightarrow{3'3\mathbf{3'3}[22'}\moverleftarrow{33']\mathbf{33'}2'2}. 
\end{gathered}
\label{eq:doubledouble}
\end{equation}
If the armlengths are equal, the $y_{ij}$ measurements from the double
bigrams shown in bold sum up to zero, and give no contribution to the GW
and secondary-noise responses; if the armlengths are different, the
presence of the bigrams introduces a preferred direction that
distinguishes the sensitivities of the 8-link and 12-link observables.
In the equal armlength limit, the observables of Eq.\ \eqref{eq:self12}
can have the same secondary-noise PSD as the standard 16-link $X_1$
observables, or they can have two new PSDs (shown as solid and dotted
curves in the bottom panel of Fig.\ \ref{fig:sens}) with nulls at $f =
k/(2L)$ and $f=(2k+1)/(4L)$, and at $f = k/(4L)$ and $f = k/(6L)$,
respectively.

At length 24 we find also some $X$-type observables derived from
\emph{non-self}-splicings of 12-link modified TDI $X$-type observables,
such as
\begin{equation}
\begin{gathered}
\moverrightarrow{3'322'22'[22'3'33'3}
\moverleftarrow{2'233'33']33'2'22'2},\\
\moverrightarrow{3'33'322'[22'22'3'3}
\moverleftarrow{2'22'233']33'33'2'2}.
\end{gathered}
\label{eq:nonselfref}
\end{equation}
These non-self-splicings occur between 12-link $L$-closed link strings
that differ by a reflection-like, noncyclic index shift (e.g., $2
\leftrightarrow 3'$, $3 \leftrightarrow 2'$, with $1$ and $1'$ reversing
time direction, if they are present in the string). In the
equal-armlength limit, these observables have again the same sensitivity
as the standard 16-link $X_1$, and they have the same secondary-noise
PSDs as either the standard 16-link $X_1$ observables or the
self-splicings of 12-link observables discussed above.

\begin{figure}
\includegraphics[width=3.0in]{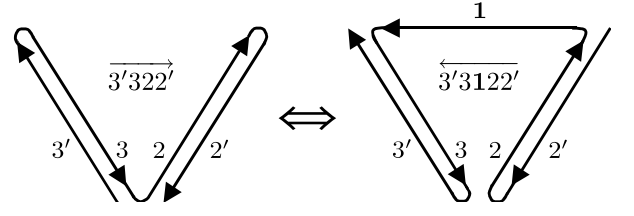}
\caption{The time-reversal plus \emph{bridge} symmetry is exploited by
certain $\dot{L}$-closed non-self-splicings at length 20 and above, such
as the link string of Eq.\
\eqref{eq:nonselfsymmetry}.\label{fig:bridge}}
\end{figure}

We now direct our consideration back to the full set of $\dot{L}$-closed
link strings of lengths 16 to 24. By exhaustive exploration, we were able to
show that all of them can be obtained as splicings of two shorter
$L$-closed link strings. An explicit specification of the splicings is
given in the lists of TDI observables available at the webpage
\url{www.vallis.org/tdi}. A good portion of the $\dot{L}$-closed link
strings are obtained from self-splicings. The non-self-splicings occur
between $L$-closed strings that are often related by evident symmetries
(such as index shifts). Interestingly, these symmetries may occur
between $L$-closed strings of different length, as it happens for the
20-link observable
\begin{equation}
\moverrightarrow{3'3[3'}|
\moverleftarrow{122'}
\moverrightarrow{1}
\moverleftarrow{3'31}
\moverrightarrow{2'213]22'}
\moverleftarrow{33'2'2},
\label{eq:nonselfsymmetry}
\end{equation}
which can be interpreted as the splicing of the 8-link $X$ string
$\moverrightarrow{3'322'}\moverleftarrow{33'2'2}$ with the reversal of
the 12-link string
$\moverleftarrow{3'3\mathbf{1}22'}\moverrightarrow{\mathbf{1}33'}
\moverleftarrow{\mathbf{1}}\moverrightarrow{2'2\mathbf{1}}$. The latter
can be obtained from the former by reversing the time direction of all
$3'3$ and $22'$ bigrams, and inserting $\overrightarrow{\mathbf{1}}$ and
$\overleftarrow{\mathbf{1}}$ \emph{bridges} as needed to keep the arrows
consecutive, as shown in Fig.\ \ref{fig:bridge}.

Also interesting is that some splicings result in $\dot{L}$-closed
strings shorter than the sum of the lengths of their components, as in
the case of the 22-link observable
\begin{equation}
\begin{gathered}
\moverrightarrow{322'21[1'132}
\moverleftarrow{1'12\mathbf{3}}]\moverrightarrow{\mathbf{3}}
\moverleftarrow{2'23}
\moverrightarrow{1'2'}
\moverleftarrow{3122'1'} \equiv \\
\moverrightarrow{322'211'132}
\moverleftarrow{1'12}
\moverleftarrow{2'23}
\moverrightarrow{1'2'}
\moverleftarrow{3122'1'},
\label{eq:shortersurprise}
\end{gathered}
\end{equation}
which can be interpreted as the splicing of the 16-link string
$\moverrightarrow{322'213}\moverleftarrow{2'23}\moverrightarrow{1'2'}
\moverleftarrow{3122'1'}$ with the 8-link $U$-type string
$\moverrightarrow{1'132}\moverleftarrow{1'123}$, occurring in such a way
that the final $\overleftarrow{3}$ of the latter is \emph{erased} by the
subsequent $\overrightarrow{3}$ from the former [both bold in Eq.\
\eqref{eq:shortersurprise}].

We conjecture that all $\dot{L}$-closed strings can be obtained as
splicings of $L$-closed strings. A possible proof could focus on the
\emph{monadic} $L$-closed strings (those that cannot be obtained as the
splicing of two $L$-closed strings), which appear at every even length:
consider for instance the string
$\moverrightarrow{3'322'22'\cdots22'}\moverleftarrow{33'2'22'2\cdots2'2}
$. To prove our conjecture, one would show that \emph{monadic}
$L$-closed strings are never $\dot{L}$-closed, as we know to be the case
up to length 24.

\section{Conclusions} \label{sec:conclusion}

I have described Geometric TDI, a powerful new approach to understand
TDI intuitively, and to interpret \emph{all} of its observables as the
measurements of virtual synthesized interferometers, extending the original 
intuition of Tinto and Armstrong \cite{tintoarm} and the graphical explanations of Shaddock \cite{stea2003} and Summers \cite{summers}. In Geometric TDI, observables consisting of $n$ one-way phase measurements are represented by \emph{link strings} of length $n$, and can be enumerated exhaustively by listing all possible link strings, and then applying simple rules to
determine which strings correspond to first-generation ($|L|$-closed),
modified ($L$-closed), or second-generation ($\dot{L}$-closed)
observables. A study of the \emph{closure symmetries} of link strings
provides clues to the general rule that modified TDI observables
combine (i.e., \emph{splice}) into second-generation observables,
maintaining in most cases the same GW sensitivity.

In addition to its pedagogical value, Geometric TDI has the practical
interest of providing a systematic method to explore the space of
second-generation TDI observables; such a method was unavailable prior
to this work. Possible applications include the optimization of GW
sensitivity (in analogy to the work of Prince and colleagues in Ref.\
\cite{optimal}), and the development of targeted noise diagnostics.

In Sec.\ \ref{sec:survey}, I have used Geometry TDI to survey all
second-generation TDI observables of lengths up to 24. My results
(available in full at the webpage \url{www.vallis.org/tdi}) show that
the garden of TDI observables contains a wealth of previously unknown
specimens; this richness only increases with string length. In Sec.\
\ref{sec:advantage}, I have pointed out how certain alternative forms of
the standard observables [such as the four-beam $X_1$ observables of
Eq.\ \eqref{eq:alternativex}] have improved GW sensitivity in realistic
measurement conditions (because they have fewer noise and GW response
nulls), and reduced susceptibility to gaps and glitches (because they
have a smaller temporal footprint).

The new observables become possible in Geometric TDI because the time
advancements and retardations of one-way phase measurements are put on
the same footing, while only retardations were considered in traditional TDI.
This generalization does not constitute a third-generation TDI,
but its combinatorial power and attractive symmetry justify its addition to
the canon of TDI. There seem to be no particular challenges to
implementing time-symmetric observables, especially in the framework of
post-processed TDI \cite{lisainterpolation}, whether performed onboard
or on the ground.

Future work on Geometric TDI should endeavor to prove the two open
conjectures formulated in this paper: namely, that all second-generation
TDI observables of any length can be generated as splicings of two
modified-TDI observables, and that there are no second-generation TDI
observables of the \emph{perfect} $\zeta$ type. (It would also be
interesting to explore a notion of \emph{relaxed} $\dot{L}$-closure that
includes the $\zeta_1$, $\zeta_2$, and $\zeta_3$ variables described by
Tinto and colleagues \cite{tea2004}.) Other promising directions of
research include the optimization of GW sensitivity (can it be
characterized geometrically?) and the exploration of \emph{generative
rules} other than splicing for second-generation TDI observables (what
is the geometric counterpart and generalization of the algebra of
observables studied in Refs.\ \cite{dhurandhar}?).

\acknowledgments I wish to thank John Armstrong, Frank Estabrook, and
Massimo Tinto for teaching me all things TDI; I thank also Daniel Shaddock and David Summers for useful discussions. I am grateful to the Supercomputing and Visualization Facility
at the Jet Propulsion Laboratory for providing CPU time on their Orion
cluster. This research was supported by the LISA Mission Science Office
at the Jet Propulsion Laboratory, Caltech, where it was performed under
contract with the National Aeronautics and Space Administration.

\appendix

\section{Rules and proofs}

\subsection{Counting rule for \texorpdfstring{$\dot{L}$-closed}{L-dot-closed} link strings}
\label{app:rule}
There is nothing magic about this rule, which is necessary and
sufficient to have a null total light-travel time at first order/degree
for \emph{generic} armlengths $L_{l}(t)$. To see why, set $t = 0$ at the beginning
of the link string; stepping through the string, increment $t$ by an
advancement $\mathit{\Gamma}_{l}(t)$ (defined as the time experienced by
light propagating along link $l$ for \emph{emission} at time $t$) for
each symbol $\moverrightarrow{l}$, and decrement $t$ by a delay
$L_{l}(t)$ for each symbol $\moverleftarrow{l}$. Each advancement
$\mathit{\Gamma}_{l}$ or delay $L_{l}$ enters the time arguments of all
subsequent $\mathit{\Gamma}_{m}$'s and $L_{m}$'s in the string,
generating terms $\{\mathit{\Gamma}_{l},-L_{l}\} \times
\{\dot{\mathit{\Gamma}}_{m},-\dot{L}_{m}\}$ in the first-order/degree
Taylor expansion of the total light-travel time. Since
$\mathit{\Gamma}_{l}(t) = L_{l}(t + \mathit{\Gamma}_{l}(t)) \simeq
L_{l}(t) + L_{l}(t) \dot{L}_l(t)$, we can replace all
$\mathit{\Gamma}_l$'s with $L_l$'s (while keeping the overall sum the
same to first order/degree) by adding a term $L_l \dot{L}_l$ for each
$\moverrightarrow{l}$ in the string. The counting rule given in the main
text is then equivalent to requiring that all the $L_{l} \dot{L}_{m}$
terms cancel one by one to yield a null total light-travel time.

\subsection{Translation rule from link strings to quasi-standard TDI
expressions}
\label{app:comprule}
\begin{enumerate}
\item Start at a bigram of type $\moverrightarrow{l}\moverleftarrow{m}$
(there must be at least one in every closed loop), shifting the string
cyclically to move the bigram close to the middle;
\item move to the left, starting with $\moverrightarrow{l}$, and write
down a $y_{ij}$ measurement for each index, according to the replacement
rules $\{1, 2, 3, 1', 2', 3'\} \equiv \{ y_{32}, y_{13}, y_{21}, y_{23},
y_{31}, y_{12} \}$; use a plus sign for ``$\overrightarrow{}$'' links
and a minus sign for ``$\overleftarrow{}$'' links;
\item while doing this, build the delay sequence to be applied to each
new $y_{ij}$, adding a retardation $;\!r$ \emph{after} having translated
each $\overrightarrow{r}$, and an advancement $;\!\overline{s}$
\emph{before} translating each $\overleftarrow{s}$;
\item after reaching the left end of the string, go back to the link
index $\overleftarrow{m}$ in the initial bigram, and move to the right,
using the same replacement rules;
\item while doing this, build the delay sequence from scratch, adding a
retardation $;\!s$ \emph{after} having translated each
$\overleftarrow{s}$, and an advancement $;\!\overline{r}$ \emph{before}
translating each $\overrightarrow{r}$.
\end{enumerate}
Shifting the string cyclically by a few positions can reduce the length
of the delay sequence.

\subsection{Proof that all self-splicings of \texorpdfstring{$L$-closed}{L-closed} observables are \texorpdfstring{$\dot{L}$-closed}{L-dot-closed}}
\label{app:closedproof}
A few lemmas are needed for this proof. Looking back to the counting of
$l\dot{m}$ pairs outlined in Sec.\ \ref{sec:combenum}, we denote as
$\mathrm{prod}[\mathrm{string}]$ the polynomial obtained by regarding
the 36 possible $l\dot{m}$ pairs as monomials, and summing them with
coefficients $\#[\moverrightarrow{l\dot{m}},\moverleftarrow{l\dot{m}}]
-\#[\moverrightarrow{l}\moverleftarrow{\dot{m}},\moverleftarrow{l}
\moverrightarrow{\dot{m}}]$. A string is $\dot{L}$-closed iff
$\mathrm{prod}[\mathrm{string}] = 0$.

\emph{Lemma 1.}---The prod of an $L$-closed string is unchanged after a
cyclic string shift. Consider the shift of a single index $m$ from one
end of the string to the other. Because the string is $L$-closed, and
must therefore have $\#[\overrightarrow{l}] = \#[\overleftarrow{l}]$,
the contribution to $\mathrm{prod}[\mathrm{string}]$ from the index $m$
sums down to zero if $m$ carries a $\overrightarrow{}$, or to $m
\dot{m}$ if it carries a $\overleftarrow{}$ (and therefore does not
multiply itself). After the shift, the contribution to
$\mathrm{prod}[\mathrm{shifted\,string}]$ from the index $m$ is the
same. Hence this lemma.

\emph{Lemma 2.}---The prods of an $L$-closed string and its reversal are
opposite numbers. This is established by noticing that
$\mathrm{prod}[\mathrm{string}] + \mathrm{prod}[\mathrm{reversal}] =
[\mathrm{string}] \times [\mathrm{string}]$ (i.e., the result of
counting one $l\dot{m}$ term, with the appropriate sign, for each $l$
with each $m$ in the string, including itself). However, for a closed
string, $[\mathrm{string}] \times [\mathrm{string}]$ must be zero; this
is because in $L$-closed strings the contribution from each index sums
down to zero, given that each index multiplies every other in equal
numbers under $\overrightarrow{}$ and $\overleftarrow{}$. Hence this
lemma.

\emph{Lemma 3.}---The cross prod $[\mathrm{string}] \times
[\mathrm{reversal}]$ of a shifted string and its shifted reversal (i.e.,
the result of counting one $l \dot{m}$ term, with the appropriate sign,
for each $l$ in the string and each $m$ in the reversal) is zero. In fact, the
cross prod is separately zero for each index in the string with all the
indices in its reversal, again because the reversal (as the original
string) is $L$-closed and has $\#[\overrightarrow{m}] =
\#[\overleftarrow{m}]$. Hence this lemma.

\emph{Lemma 4.}---All self-splicings can be brought into a normal form
given by the concatenation (``$|$'') of the shifted string and its
shifted reversal, for appropriate shifts. This does not change the prod
of the self-splicing.

\emph{Proof.}---Hence, $\mathrm{prod}[\mbox{self-splicing}]$ is given by
\begin{equation}
\begin{aligned}
& \mathrm{prod}[\mathrm{shifted\,string}|\mathrm{shifted\,reversal}] = \\
& \quad \quad \phantom{\,+\,} \mathrm{prod}[\mathrm{string}] +
\mathrm{prod}[\mathrm{reversal}] \\
& \quad \quad + [\mathrm{shifted\,string}] \times [\mathrm{shifted\,reversal}]
 = 0,
\end{aligned}
\end{equation}
because the first two terms are opposite numbers and the third term vanishes. Hence the proof.

\subsection{Algebraic expressions for all second-generation TDI
observables of length 16}
\label{app:secondsixteen}

In this section I give explicit algebraic expressions for all the
second-generation TDI observables of length 16, as found in my
exhaustive survey, modulus the symmetries discussed in Sec.\
\ref{sec:combenum}. There is considerable arbitrariness in writing these
expressions, corresponding to the selection of representative link
strings in each equivalence class, to the convention used in translating
strings to sums of $y_{ij}$ measurements, and to the choice of the
initial time of evaluation for each observable. Here I list link strings
in a \emph{normal form} whereby each string begins with the largest
continuous substring of forward-time indices; I adopt the translation
rules given in Sec.\ \ref{sec:combenum} [just below Eq.\
\eqref{eq:x1splicing}]; and I adjust the time of evaluation to minimize
the length of the longest delay sequence in the expression.

\renewcommand{\o}[1]{\overline{#1}}
\doublemylistitem{X type.}
The standard 16-link observable $X_1^{16,2}$ is obtained by applying the
translation rules to the link string
$\moverrightarrow{3'322'22'3'3}\moverleftarrow{2'233'33'2'2}$, and
evaluating the resulting expression at the initial time
$t_{;3'322'22'3'3}$:
\begin{equation}
\begin{aligned}
 \,&  y_{132;3 2 2'2 2'3'3 } 
+ y_{231;2 2'2 2'3'3 } 
+ y_{123;2'2 2'3'3 } \\
+\,& y_{321;2 2'3'3 } 
+ y_{123;2'3'3 } 
+ y_{321;3'3 } 
+ y_{132;3} 
+ y_{231} \\
-\,& y_{32'1} 
- y_{123;2'} 
- y_{231;2 2'} 
- y_{132;3 2 2'}
- y_{231;3'3 2 2'} \\
-\,& y_{132;3 3'3 2 2'}
- y_{321;3'3 3'3 2 2'} 
- y_{123;2'3'3 3'3 2 2'};
\end{aligned}
\end{equation}
the alternative 16-link observables $X_1^{16,4,\pm 1}$ can be written
from the link strings
$\moverrightarrow{3'322'3'3}\moverleftarrow{2'233'}\moverrightarrow{22'}
\moverleftarrow{33'2'2}$ and
$\moverrightarrow{22'3'322'}\moverleftarrow{33'2'2}\moverrightarrow{3'3}
\moverleftarrow{2'233'}$, evaluated at times $t_{3'322'3'3\o{2'2}}$ and
$t_{22'3'322'\o{33'}}$, respectively:
\begin{equation}
\begin{aligned}
 \,& y_{132;3 2 2'3'3 \o{2'2}}
+ y_{231;2 2'3'3 \o{2'2}} 
+ y_{123;2'3'3 \o{2'2}} \\
+\,& y_{321;3'3 \o{2'2}} 
+ y_{132;3 \o{2'2}} 
+ y_{231;\o{2'2}} 
- y_{321;\o{2'2}} 
- y_{123;\o{2}} \\ 
-\,& y_{231} 
- y_{132;3 } 
+ y_{123;\o{2}3'3 } 
+ y_{321;\o{2'2}3'3 } 
- y_{231;\o{2'2}3'3} \\
-\,& y_{132;3 \o{2'2}3'3 } 
- y_{321;3'3 \o{2'2}3'3 } 
- y_{123;2'3'3 \o{2'2}3'3},
\end{aligned}
\end{equation}
\begin{equation}
\begin{aligned}
 \,& y_{123;2'3'3 2 2'\o{33'}} 
+ y_{321;3'3 2 2'\o{33'}}
+ y_{132;3 2 2'\o{33'}} \\
+\,& y_{231;2 2'\o{33'}}
+ y_{123;2'\o{33'}}
+ y_{321;\o{33'}}
- y_{231;\o{33'}} 
- y_{132;\o{3'}} \\ 
-\,& y_{32'1}
- y_{123;2'} 
+ y_{132;\o{3'}2 2'} 
+ y_{231;\o{33'}2 2'}
- y_{321;\o{33'}2 2'} \\
-\,& y_{123;2'\o{33'}2 2'} 
- y_{231;2 2'\o{33'}2 2'}
- y_{132;3 2 2'\o{33'}2 2'};
\end{aligned}
\end{equation}
last, the null 16-link observable $X_1^{16,4,0}$ can be written from the
link string
$\moverrightarrow{22'22'}\moverleftarrow{33'2'2}\moverrightarrow{3'33'3}
\moverleftarrow{2'233'}$, evaluated at time $t_{22'22'\o{33'2'2}}$:
\begin{equation}
\begin{aligned}
 \,& y_{123;2'2 2'\o{33'2'2}}
+ y_{321;2 2'\o{33'2'2}}
+ y_{123;2'\o{33'2'2}} \\
+\,& y_{321;\o{33'2'2}}
- y_{231;\o{33'2'2}}
- y_{132;\o{3'2'2}}
- y_{321;\o{2'2}} 
- y_{123;\o{2}} \\
+\,& y_{132;\o{3'}}
+ y_{231;\o{33'}}
+ y_{132;\o{3'33'}}
+ y_{231;\o{33'33'}}
- y_{321;\o{33'33'}} \\
-\,& y_{123;2'\o{33'33'}}
- y_{231;2 2'\o{33'33'}}
- y_{132;3 2 2'\o{33'33'}}.
\end{aligned}
\end{equation} 
Expressions for the $X$-type observables that use the arms $1$, $2$ and
$1$, $3$ are obtained by cyclic index shifts.

\doublemylistitem{U type.}
The three 16-link $U$-type observables
that use the oriented arms $1$, $1'$, $2$, and $3$
correspond to the link strings
\begin{equation}
\begin{gathered}
\moverrightarrow{3211'}\moverleftarrow{23123}
\moverrightarrow{1'132}\moverleftarrow{1'11'}, \\
\moverrightarrow{3211'}\moverleftarrow{23}
\moverrightarrow{1'132}\moverleftarrow{1'12311'}, \\
\moverrightarrow{3211'132}\moverleftarrow{1'123}
\moverrightarrow{1'}\moverleftarrow{2311'};
\end{gathered}
\end{equation}
applying the translation rules and evaluating at the times
$t_{3211'\o{2312}}$, $t_{3211'\o{23}1'1}$, and $t_{3211'132\o{1}}$,
respectively, yields
\begin{equation}
\begin{aligned}
\,& y_{231;2 1 1'\o{2312}}
+ y_{123;11'\o{2312}}
+ y_{312;1'\o{2312}} \\
+\,& y_{213;\o{2312}}
- y_{123;\o{2312}}
- y_{231;\o{312}}
- y_{312;\o{12}}
- y_{123;\o{2}} \\
-\,& y_{231}
+ y_{213;\o{1'}3 }
+ y_{312;\o{11'}3 }
+ y_{231;\o{311'}3}
+ y_{123;\o{2311'}3 } \\
-\,& y_{213;\o{2311'}3}
- y_{312;1'\o{2311'}3 }
- y_{213;1 1'\o{2311'}3 },
\end{aligned}
\end{equation}
\begin{equation}
\begin{aligned}
\,& y_{231;2 1 1'\o{23}1'1 }
+ y_{123;1 1'\o{23}1'1 }
+ y_{312;1'\o{23}1'1} \\
+\,& y_{213;\o{23}1'1 }
- y_{123;\o{23}1'1 }
- y_{231;\o{3}1'1}
+ y_{213;1}
+ y_{312} \\
+\,& y_{231;\o{3}}
+ y_{123;\o{23}}
- y_{213;\o{23}}
- y_{312;1'\o{23}}
- y_{123;1 1'\o{23}} \\
-\,& y_{231;2 1 1'\o{23}}
- y_{312;3 2 11'\o{23}}
- y_{213;1 3 2 1 1'\o{23}},
\end{aligned}
\end{equation}
\begin{equation}
\begin{aligned}
\,& y_{231;2 1 1'1 3 2 \o{1'}}
+ y_{123;1 1'1 3 2 \o{1'}}
+ y_{312;1'1 3 2 \o{1'}} \\
+\,& y_{213;1 3 2 \o{1'}}
+ y_{312;3 2 \o{1'}}
+ y_{231;2 \o{1'}}
+ y_{123;\o{1'}}
- y_{213;\o{1'}} \\
-\,& y_{312}
- y_{123;1 }
- y_{231;2 1}
+ y_{213;\o{1'}3 2 1 }
- y_{123;\o{1'}3 2 1 } \\
-\,& y_{231;2 \o{1'}3 2 1}
- y_{312;3 2 \o{1'}3 2 1 }
- y_{213;1 3 2 \o{1'}3 2 1 };
\end{aligned}
\end{equation}
the third expression is closest to the $U_1$ given in Ref.\
\cite{tea2004} (but see note \ref{notetea}). Expressions for the
$U$-type observables that other sets of oriented arms (i.e.,
$\{2,2',1,3\}$, $\{3,3',1,2\}$, $\{1,1',2',3'\}$, $\{2,2',1',3'\}$, and
$\{3,3',1',2'\}$) are obtained by cyclic and noncylic index shifts.

\doublemylistitem{E type.} The three 16-link $E$-type observables that
use the oriented arms $1$, $1'$, $2'$, and $3$ correspond to the link
strings
\begin{equation}
\begin{gathered}
\moverrightarrow{11'13}\moverleftarrow{2'1'1}
\moverrightarrow{2'}\moverleftarrow{3}
\moverrightarrow{1'2'}\moverleftarrow{311'}
\moverrightarrow{3}\moverleftarrow{2'}, \\
\moverrightarrow{1'11'2'}\moverleftarrow{311'}
\moverrightarrow{3}\moverleftarrow{2'}
\moverrightarrow{13}\moverleftarrow{2'1'1}
\moverrightarrow{2'}\moverleftarrow{3}, \\
\moverrightarrow{11'2'}\moverleftarrow{3}
\moverrightarrow{1'13}\moverleftarrow{2'1'1}
\moverrightarrow{2'}\moverleftarrow{311'}
\moverrightarrow{3}\moverleftarrow{2'};
\end{gathered}
\end{equation}
applying the translation rules and evaluating at the times
$t_{11'13\o{2'1'1}2'}$, $t_{1'11'2'\o{311'}3}$, and
$t_{11'2'\o{3}1'13\o{2'}}$, respectively, yields:
\begin{equation}
\begin{aligned}
\,& y_{312;1'1 3 \o{2'1'1}2'}
+ y_{213;1 3 \o{2'1'1}2'}
+ y_{312;3 \o{2'1'1}2'} \\
+\,& y_{231; \o{2'1'1}2'}
- y_{321; \o{2'1'1}2'}
- y_{213; \o{1'1}2'}
- y_{312; \o{1}2'}
+ y_{32'1} \\
-\,& y_{231}
+ y_{213; \o{1'}3}
+ y_{321; \o{2'1'}3 }
- y_{231; \o{2'1'}3 }
- y_{312;3 \o{2'1'}3} \\
-\,& y_{213;1 3 \o{2'1'}3 }
+ y_{231; \o{3}1'1 3 \o{2'1'}3}
- y_{321; \o{3}1'1 3 \o{2'1'}3 },
\end{aligned}
\end{equation}
\begin{equation}
\begin{aligned}
\,& y_{213;1 1'2'\o{311'}3 }
+ y_{312;1'2'\o{311'}3}
+ y_{213;2'\o{311'}3 } \\
+\,& y_{321;\o{311'}3}
- y_{231;\o{311'}3 }
- y_{312;\o{11'}3 }
- y_{213;\o{1'}3}
+ y_{231} \\
-\,& y_{32'1}
+ y_{312;\o{1}2'}
+ y_{231;\o{31}2'}
- y_{321;\o{31}2'}
- y_{213;2'\o{31}2'} \\
-\,& y_{312;1'2'\o{31}2'}
+ y_{321;\o{2'}1 1'2'\o{31}2'}
- y_{231;\o{2'}1 1'2'\o{31}2'},
\end{aligned}
\end{equation}
\begin{equation}
\begin{aligned}
\,& y_{312;1'2'\o{3}1'1 3 \o{2'}}
+ y_{213;2'\o{3}1'1 3 \o{2'}}
+ y_{321;\o{3}1'1 3 \o{2'}} \\
-\,& y_{231;\o{3}1'1 3 \o{2'}}
+ y_{213;1 3 \o{2'}}
+ y_{312;3 \o{2'}}
+ y_{231;\o{2'}}
- y_{321;\o{2'}} \\
-\,& y_{21'3}
- y_{312;1'}
+ y_{321;\o{2'}1 1'}
- y_{231;\o{2'}1 1'}
- y_{312;3 \o{2'}1 1'} \\
-\,& y_{213;1 3 \o{2'}1 1'}
+ y_{231;\o{3}1'1 3 \o{2'}1 1'}
- y_{321;\o{3}1'1 3 \o{2'}1 1'};
\end{aligned}
\end{equation}
the third expression is closest to the $E_1$ given in Ref.\
\cite{tea2004}. Expressions for the other possible sets of oriented arms
(i.e., $\{2,2',1,3'\}$ and $\{3,3',1',2\}$) are obtained by cyclic index
shifts.

\doublemylistitem{P type.}
The three 16-link $P$-type observables that use the oriented arms $1$,
$1'$, $2$, and $3'$ correspond to the link strings
\begin{equation}
\begin{gathered}
\moverrightarrow{3'1'11'}\moverleftarrow{2}
\moverrightarrow{3'}\moverleftarrow{11'3'}
\moverrightarrow{21}\moverleftarrow{3'}
\moverrightarrow{2}\moverleftarrow{1'12}, \\ 
\moverrightarrow{211'1}\moverleftarrow{3'}
\moverrightarrow{2}\moverleftarrow{1'12}
\moverrightarrow{3'1'}\moverleftarrow{2}
\moverrightarrow{3'}\moverleftarrow{11'3'}, \\
\moverrightarrow{211'}\moverleftarrow{2}
\moverrightarrow{3'1'1}\moverleftarrow{3'}
\moverrightarrow{2}\moverleftarrow{1'12}
\moverrightarrow{3'}\moverleftarrow{11'3'}; 
\end{gathered}
\end{equation}
applying the translation rules and evaluating at the times
$t_{3'1'11'\o{2}3'\o{11'}}$,
$t_{211'1\o{3'}2\o{1'1}}$, and
$t_{211'\o{2}3'1'1\o{3'}}$,
respectively, yields
\begin{equation}
\begin{aligned}
\,& y_{132;1'1 1'\o{2}3'\o{11'}}
+ y_{213;11'\o{2}3'\o{11'}}
+ y_{312;1'\o{2}3'\o{11'}} \\
+\,& y_{213;\o{2}3'\o{11'}}
- y_{123;\o{2}3'\o{11'}}
+ y_{132;\o{11'}}
- y_{312;\o{11'}}
- y_{213;\o{1'}} \\
-\,& y_{13'2}
+ y_{123;\o{2}3'}
+ y_{312;\o{12}3'}
- y_{132;\o{12}3'}
+ y_{123;\o{2}3'\o{12}3'} \\
-\,& y_{213;\o{2}3'\o{12}3'}
- y_{312;1'\o{2}3'\o{12}3'}
- y_{123;1 1'\o{2}3'\o{12}3'},
\end{aligned}
\end{equation}
\begin{equation}
\begin{aligned}
\,& y_{123;1 1'1 \o{3'}2 \o{1'1}}
+ y_{312;1'1 \o{3'}2 \o{1'1}}
+ y_{213;1\o{3'}2 \o{1'1}} \\
+\,& y_{312;\o{3'}2 \o{1'1}}
- y_{132;\o{3'}2 \o{1'1}}
+ y_{123;\o{1'1}}
- y_{213;\o{1'1}}
- y_{312;\o{1}} \\
-\,& y_{123}
+ y_{132;\o{3'}2 }
+ y_{213;\o{1'3'}2 }
- y_{123;\o{1'3'}2}
+ y_{132;\o{3'}2 \o{1'3'}2 } \\
-\,& y_{312;\o{3'}2 \o{1'3'}2}
- y_{213;1 \o{3'}2 \o{1'3'}2 }
- y_{132;1'1 \o{3'}2 \o{1'3'}2},
\end{aligned}
\end{equation}
\begin{equation}
\begin{aligned}
\,& y_{123;1 1'\o{2}3'1'1 \o{3'}}
+ y_{312;1'\o{2}3'1'1 \o{3'}}
+ y_{213;\o{2}3'1'1\o{3'}} \\
-\,& y_{123;\o{2}3'1'1 \o{3'}}
+ y_{132;1'1 \o{3'}}
+ y_{213;1\o{3'}}
+ y_{312;\o{3'}}
- y_{132;\o{3'}} \\
+\,& y_{123;\o{2}}
- y_{213;\o{2}}
- y_{312;1'\o{2}}
- y_{123;1 1'\o{2}}
+ y_{132;\o{3'}2 11'\o{2}} \\
-\,& y_{312;\o{3'}2 1 1'\o{2}}
- y_{213;1 \o{3'}2 11'\o{2}}
- y_{132;1'1 \o{3'}2 1 1'\o{2}};
\end{aligned}
\end{equation}
the third expression is closest to the $P_1$ given in Ref.\
\cite{tea2004}. Expressions for the other possible sets of oriented arms
(i.e., $\{2,2',1',3\}$ and $\{3,3',1,2'\}$) are obtained by cyclic index
shifts.
\lastlistitem



\end{document}